\documentclass[pra,eqsecnum,twocolumn,amsfonts]{revtex4}
\usepackage{amsfonts}
\usepackage{amsmath}
\usepackage{bm}
\usepackage{epsf}
\usepackage{graphicx}

\newcommand{\beq}{\begin{equation}}
\newcommand{\eeq}{\end{equation}}
\newcommand{\beqa}{\begin{eqnarray}}
\newcommand{\eeqa}{\end{eqnarray}}

\newcommand{\ket}[1]{| #1    \rangle }

\newcommand{\ave}[1]{  \langle #1   \rangle }
\newcommand{\qave}[2]{  \langle #1  | #2  | #1  \rangle }

\newcommand{\rref}[1]{~(\ref{#1})}

\begin{document}


\newcommand{\Qsigma}{g_0}
\newcommand{\covQ}{G}
\newcommand{\covP}{H}
\newcommand{\corQ}[1]{g_{#1}}
\newcommand{\corP}[1]{h_{#1}}

\newcommand{\suba}[1]{_{_{A_{#1}}}}
\newcommand{\subat}[1]{_{_{\widetilde{A}_{#1}}}}
\newcommand{\subb}[1]{_{_{B_{#1}}}}
\newcommand{\subbt}[1]{_{_{\widetilde{B}_{#1}}}}
\newcommand{\subab}[1]{_{_{A_{#1}B_{#1}}}}
\newcommand{\subabt}[1]{_{_{\widetilde{A}_{#1}\widetilde{B}_{#1}}}}


\title
{
Spatial structures and localization of vacuum entanglement \\ in
the linear harmonic chain
}

\author{ Alonso Botero }
\email{abotero@uniandes.edu.co} \affiliation{
    Departamento de F\'{\i}sica,
    Universidad de Los Andes,
    Apartado A\'ereo 4976,
    Bogot\'a, Colombia.}
\author{ Benni Reznik }
\email{reznik@post.tau.ac.il} \affiliation{ Department of Physics
and Astronomy, Beverly and Raymond Sackler Faculty of Exact
Sciences, Tel-Aviv University, Tel Aviv 69978, Israel.
       }

\date{\today}

\begin{abstract}
\bigskip
We study the structure of vacuum  entanglement for two
complimentary segments of a linear harmonic chain, applying the
modewise decomposition of entangled gaussian states discussed in
\cite {modewise}. We find that the resulting entangled mode shape
hierarchy shows a distinctive layered structure with well defined
relations between  the depth of the modes, their characteristic
wavelength, and their entanglement contribution. We re-derive in
the strong coupling (diverging correlation length) regime, the
logarithmic dependence of entanglement on the segment size
predicted by conformal field theory for the boson universality
class, and discuss its relation with the mode structure. We
conjecture that the persistence of vacuum entanglement between
arbitrarily separated finite size regions is connected with the
localization of the highest frequency innermost modes.
\end{abstract}
 \maketitle
\section{Introduction}

The study of entanglement properties of a number of physical
models including spin chains, coupled fermions  and harmonic
oscillators [1-10], has revealed interesting aspects of
entanglement in spatially extended many body systems.

On one hand, a number of universal aspects connected to the
behavior of the two-point correlation function has been verified.
For instance, it has been shown that one-dimensional XY and
Heisenberg spin chains near a critical point \cite{vidal}, leads
to the same scaling behavior for massless boson and fermion
universality classes as predicted by conformal field theory
\cite{callan,kabat,holzhey}. In the massless case, the
entanglement between a region of size $L$ and the remainder of the
system, increases either as $\sim \frac{1}{3}\log L$ (bosons) or
$\sim \frac{1}{6}\log L$ (fermions). These characteristic
behaviors relate to a one-dimensional version of the black hole
entropy ``area-law" \cite{bombelli,srednicki,dowker}.

In contrast to these universal results,  other aspects of many
body entanglement have proven to be model dependent and not
entirely captured by the behavior of the correlation functions.
For example, the entanglement length defined in \cite{EL1} as an
analog of the correlation length, has been shown for a family of
models be infinite while the correlation length is finite
\cite{EL2,pachos}. The converse situation has also been
demonstrated for spin chains \cite{nielsen,osterloh} and harmonic
chains\cite{unpublished}. In these cases, while the correlation
length  can be infinite or large, the entanglement between two
sites truncates to zero for a relatively small separation. It is
therefore fair to say that we still lack a generic
characterization of  entanglement in many body systems.

In this paper we concentrate on an aspect that has
 received relatively little attention, but which could potentially
  sharpen the emerging picture. This is the
\emph{entanglement structure} dictated by many-body interactions--the connection
between the form of the Hamiltonian, the
quantum state entanglement structure, and the spacial
distributions  associated with the quantum states.
Our aim is to provide a detailed
analysis of the  bipartite ground state entanglement
structure of the linear harmonic chain.

In general the determination of the entanglement structure of a
given quantum state is a complicated problem; however, things are
greatly simplified for the vacuum state of an oscillator chain,
which is a pure state of the Gaussian family \cite{gaussian}. An
important feature of multi-mode pure Gaussian states is their
fundamentally simple structure with respect to bipartite
entanglement: it can be shown that pure Gaussian state
entanglement is equivalent to products of entangled pairs of
single modes \cite{modewise,cirac}, so that the total entanglement
is the sum of the $1\times 1$ modewise entanglement contributions;
therefore, the canonical structure of gaussian bipartite
entanglement is
 $1\times 1$-mode gaussian
entanglement. We therefore aim  to characterize the spatial
structure of the entangled modes within each of two complementary
regions of the harmonic chain  and connect  this spacial structure
to the corresponding  entanglement contributions. This analysis is
performed for both the weak and strong coupling regimes.

An extrapolation of our  results to the continuum limit shows
agreement with previously known results, such as the
$\frac{1}{3}\ln L$ entanglement behavior for bosons, as well as
provides new insight into the entanglement characteristics of the
vacuum. In particular, it shows that the inclusion of an
ultraviolet cutoff--in any way necessary to regulate the
corresponding massless relativistic quantum field theory in the
presence of interactions--gives rise to a localization of the
highest frequency modes around the midpoint of the block. Thus, it
appears that contrary to the behavior of correlations,  the
long-distance behavior of entanglement between localized regions
is directly connected with the high frequency modes. This somewhat
paradoxical feature of field entanglement was noticed earlier in
connection with the extraction of entanglement from the vacuum
\cite{vacent} and violation of Bell's inequalities in the vacuum
\cite{vac-bell-ineq}.

The article is organized as follows. In Section II. we begin with
a short review of Gaussian states and the mode-wise structure of
entanglement. We then study the structure of the local mode shapes
within a block and their form by introducing a mode
``participation function''. In Section III. we specialize to the
case of the linear harmonic chain ground state, and study in some
detail the dependence of the correlation functions on the coupling
strength and size of the chain. We show that three regimes of
behavior can be identified for a chain of fixed size and varying
coupling strength. In Section IV. we present a simple application
of the mode mapping for the case of a single oscillator with
respect to the rest of the chain, and demonstrate how in this
simple case, the properties of the correlation function show up in
the behavior of the entanglement entropy. Section V surveys the
main results of this paper on a qualitative level based on a
numerical analysis of the modewise structure in the weak and
strong coupling limits. This survey is complemented in section VI
with an analytic  study of the spatial modewise structure as well
as the derivation of the $\frac{1}{3}\ln L$ ``area" law. Our
results are compared with previous related work in Section VII.
Finally we conclude in Section VIII.  In the Appendix we provide a
connection between the discrete and continuum correlation
functions.

\section{Gaussian State Modewise Entanglement}
\label{gaussect}

The ground state of a linear chain of oscillators is a pure
Gaussian state \cite{gaussian}. Gaussian entanglement is
characterized by the following simplifying property
\cite{modewise} (see also \cite{cirac} for an alternative proof):
If $\ket{\psi}_{AB}$ is any Gaussian pure state of  $N$ modes
entangling modes in two regions $A$ and modes in $B$, then
$\ket{\psi}$ may always be written as a product of two-mode and
one-mode states
\begin{equation}\label{statedecomp}
\ket{\psi}_{AB} =
\ket{\widetilde{\psi}_1}\subabt{1}\ket{\widetilde{\psi}_2}\subabt{2}\ldots
\ket{\widetilde{\psi}_s}\subabt{s}\ket{0}\subat{F}\ket{0}\subbt{F}
\end{equation}
where the $\ket{\widetilde{\psi}_i}\subabt{i}$ are entangled
states of  one mode from the set $A$ and one mode from the set $B$
and $\ket{0}\subat{F}$ and $\ket{0}\subbt{F}$ are products of
vacuum states for the remaining modes.  The fact that any pure
Gaussian state can be decomposed according to \rref{statedecomp}
implies that the bi-partite entanglement of the  state is the sum
of the entanglements from each one of the participating pairs. In
turn, the entanglement of each pair is the von-Neumann entropy of
the reduced density matrix  obtained from the pairwise state
$\ket{\widetilde{\psi}_i}\subabt{i}$. In addition to the
quantification of the total amount of entanglement, it also
becomes relevant to investigate the relative contribution of the
individual entangled modes, together with the relationship that
may exist between the ``shape" of these modes and their
entanglement contribution. We use this section therefore to review
and introduce some general technical aspects of pure entangled
Gaussian state analysis.

\subsection{Review of Gaussian States}

We begin by  reviewing some basic facts about Gaussian (pure or
mixed) states. Let us represent the canonical variables of an
$N$-mode system by the vector
\begin{equation}
    \eta = ( q, p )^T \, ,
\end{equation}
where $q = (q_1, q_2, ..., q_N)^T$, $p = (p_1, p_2, ..., p_N)^T$.
The commutation relations may thus be expressed as
\begin{equation}
[\eta_\alpha,\eta_\beta] = i J_{\alpha \beta}
\end{equation}
where $J$ is the so-called \emph{symplectic matrix}
\begin{equation}
J = \left(%
\begin{array}{cc}
  0 & \openone \\
  -\openone & 0 \\
\end{array}%
\right)\, .
\end{equation}
A Gaussian quantum state $\rho$ for a set of $N$ modes is uniquely
characterized by the first and second moments of $\eta$. In
dealing with entanglement aspects of Gaussian states, a shift in
the expectation value of the canonical variables corresponds to a
local operation. Thus, it may be assumed throughout that  that
$\ave{\eta}=0$. In such a case,  the state is entirely
characterized by the  matrix of second moments, the so-called
phase-space $2N\times 2N$ covariance matrix (CM):
\begin{equation}\label{covmat}
M = \mathrm{Re}\ave{\eta \, \eta^T} \, .
\end{equation}
Of particular interest is the group of transformations preserving
the Gaussian character of the state. Within the family of states
with $\ave{\eta}=0$, the group is   the homogeneous group of
linear symplectic transformations $S \in Sp(2N,R)$ preserving the
commutation relations under $\tilde\eta = S \eta$, or
equivalently, preserving the symplectic matrix under the
similarity transformation $ S J S^T =  J\, $. Under a symplectic
transformation, a Gaussian state characterized by a covariance
matrix $M$ gets mapped to a Gaussian state characterized by the
covariance matrix $\tilde M = S M S^T$.

Somewhat analogous to the construction of normal modes for a
linear system  is the construction of modes in which the
covariance matrix takes a particularly simple form. A theorem due
to Williamson\cite{williamson,will-simon} states that a certain
symplectic transformation $S_W$ always exists that brings $M$ to
the normal form (``Williamson normal form")
\begin{equation}
W = S_W M S_W^T = \mathrm{diag}(\lambda_1,
\lambda_2,...\lambda_N,\lambda_1, \lambda_2,...\lambda_N) ,
\end{equation}
where the diagonal elements $\lambda_1, \lambda_2,...\lambda_N$
are referred to as the symplectic eigenvalues and must be greater
than or equal to $1/2$ according to the uncertainty principle. The
transformation $S_W$ defines a new set of modes (``Williamson
modes") $\tilde{\eta}{(m)}$, with corresponding annihilation
operators $ \tilde a_m = \frac{\tilde q_m + i \tilde p_m
}{\sqrt{2}} $. In terms of these modes, the state $\rho$ may be
written as a product of oscillator ``thermal" states
\cite{simon94}
\begin{equation}\label{rhogauss}
\rho = \bigotimes_m (1-e^{-\beta_m})e^{ - \beta_m\tilde{n}_m} \, ,
\end{equation}
where $\tilde{n}_m = \tilde a_{m}^\dag \tilde a_{m}$ is the
Williamson number operator associated with the
creation/annihilation operators. The average number operator obeys
a Bose-Einstein distribution
\begin{equation}
\ave{\tilde n_m} = {1\over e^{\beta_m} -1} \, .
\end{equation}
Since the covariance matrix is now diagonal with $\ave{(q_m)^2}=
\ave{(p_m)^2} =\ave{\tilde n_m} + {1\over2} $, the symplectic
eigenvalues can be related to the average number operator
according to
\begin{equation}
\lambda_m =\ave{\tilde n_m} + {1\over2} \, ,
\end{equation}
 and connected with the Boltzmann factor and $\beta_m$
 by the relations
\begin{equation}
\beta_m = \ln \frac{ \lambda_m+1/2}{\lambda_m -1/2} \, ,
\end{equation}
or
\begin{equation}     \label{sim-beta}
\lambda_m = {1\over 2} \coth\left({{1\over2}\beta_m}\right) \, .
\end{equation}
In the case of $\lambda_m = 1/2$, the thermal state reduces to a
projector onto the vacuum state $\ket{0}_m$ annihilated by the
destruction operator $\tilde a_{w_i}$. Otherwise, the state is a
mixed state, with a von Neumann entropy $S(\lambda_m)$, where
 (using natural units)
\begin{equation}\label{entropy}
S(\lambda) = \left(\lambda\! +\! 1\!/2\right)\ln\left(\lambda\!
+\! 1\!/2\right) -\left(\lambda \! -\!
1\!/2\right)\ln\left(\lambda\! -\!1\!/2\right) \, .
\end{equation}

\subsection{Modewise Entanglement}

Turning now to the entanglement of pure Gaussian states, suppose
the $N$ modes are partitioned  into two sets $\eta_A$ and
$\eta_B$. Then, the particular
 set of modes in terms of which the decomposition
\ref{statedecomp} is achieved  are those which, under local
symplectic transformations bring the local covariance matrices
$M_A = \mathrm{Re} \ave {\eta_A \eta_A^T }$ and $M_B = \mathrm{Re}
\ave {\eta_B \eta_B^T }$ into Williamson normal form. The
decomposition is related with a general property of covariance
matrices which satisfy the ``isotropic condition"
\begin{equation}\label{isotropic}
-(J M)^2 = \lambda_o^2 \openone \, ,
\end{equation}
a condition that is satisfied by the covariance matrix of any pure
Gaussian state with $\lambda_o = 1/2$. Partitioning the vector of
all the modes as $\eta = \eta_A \oplus \eta_B$, the full
covariance matrix of the pure state  may be written in block form
as
\begin{equation}
M=\mathrm{Re}\ave{\eta\, \eta^T}=\left(%
\begin{array}{cc}
M_A & K \\
K^T & M_B \\
\end{array}%
\right)\, , \
\end{equation}
where $K = \ave{\eta_A \eta_B}$. It is then possible to show
\cite{modewise} that as a consequence of the isotropic condition,
$M_A$ and $M_B$ share the \emph{same} symplectic spectrum in the
respective sectors where the symplectic eigenvalues are larger
than $1/2$. By performing local symplectic transformations
$\widetilde{\eta}_A = S_A \eta_A$ and $\widetilde{\eta}_B = S_B
\eta_B$ bringing $M_A$ and $M_B$ to Williamson normal form, it is
then possible to show that the obtained correlation matrix
$\widetilde{K} = S_A K S_B^T$ connects only the sectors  in $A$
and  $B$ with the same symplectic eigenvalue and vanishes on the
elements with symplectic eigenvalue $1/2$.  This means that if the
local symplectic spectrum is not degenerate (apart from the sector
with symplectic eigenvalue 1/2), the transformed correlation
matrix $\widetilde{K}$ connects only those pairs of modes in $A$
and $B$ with the same local symplectic eigenvalue. On the other
hand, if there are  degeneracies  in the local symplectic
spectrum,   one can still perform an additional one-sided local
orthogonal symplectic transformation that brings $\widetilde{K}$
to a form connecting the degenerate modes in a pairwise fashion.

In this paper, we will only be dealing with Gaussian states for
which the correlations between the $q$'s and the $p$'s vanish.
This affords a particular simplification in the investigation of
the entangled mode structure. For this we introduce the following
notation for the coordinate and momentum covariance matrices:
\begin{equation}
G = \ave{q q^T}\, ,  \ \ \ \ H = \ave{q q^T} \, .
\end{equation}
In the absence of $q-p$ correlations, the local covariance
matrices may thus be written in block diagonal form as
\begin{equation}\label{formma}
M_A=\mathrm{Re}\ave{\eta_A\, \eta_A^T}=\left(%
\begin{array}{cc}
G_A & 0 \\
0 & H_A \\
\end{array}%
\right)\, . \
\end{equation}
The local symplectic spectrum can then be obtained from the square
root of the doubly degenerate spectrum of the matrix $-(J_A
M_A)^2$ (where $J_A$ is the symplectic matrix of the A-modes), a
matrix which is
\begin{equation}
-(J_A M_A)^2 = \left(\begin{array}{cc}
H_A G_A & 0 \\
0 & G_A H_A \\
\end{array}
\right)\, . \
\end{equation}
Thus, the symplectic eigenvalues are given by the square root of
the eigenvalues of $H_A G_A$ (or $ G_A H_A$). Therefore we can
express a certain two-mode state in the decomposition
(\ref{statedecomp}) as \beq \ket{\widetilde{\psi}_m}_{\tilde
A_m\tilde B_m} = \sqrt{1-e^{-\beta_m}} \, \sum_n e^{-\beta_m n/2}
\ket{\tilde n}_{\tilde A_m}\ket{ m}_{\tilde B_m} \, . \eeq The
total bi-partite entanglement is then given by the sum of two-mode
contributions $\sum_{\lambda_m} S(\lambda_m)$.

\subsection{Mode Shapes}

The number states $\ket{\tilde n}_{\tilde A_m}$ in the above
two-mode Gaussian state, are eigenstates  of the Williamson modes
number operator $\tilde N_{A_m}$, which in turn can be expressed
as a combination of the local modes $\tilde\eta_{A}$. The question
is then how do the local modes contribute to each of the
collective Williamson modes? This can be answered by studying the
symplectic transformation $\widetilde{\eta}_A = S_A \eta_A $
between the local and global modes. Particularly, we now wish to
directly relate the symplectic transformation to the eigenvectors
of $H_A G_A$ and $G_A H_A$. First note that in the absence of
$q-p$ correlations, the symplectic transformation will not mix the
$q$'s and $p$'s, so we may write it as
\begin{equation}
\tilde{q}_A = X q_A  \, \ \ \ \ \tilde{p}_A = Y p_A    \, \ \ \ X
Y^T = \openone
\end{equation}
where the last condition guarantees that the transformation is
symplectic. Now define $\Lambda$ to be the diagonal matrix with
the symplectic eigenvalues of $M_A$. Thus $\Lambda$ may be written
as
\begin{equation}\label{transGH}
\Lambda = X G_A X^T = Y H_A Y^T \, .
\end{equation}
Now, let $u^{(m)}$ and $v^{(m)} $ be  right eigenvectors of $ H_A
G_A$ and $  G_A H_A$ respectively corresponding to the symplectic
eigenvalue (the $m$-th entry in $\Lambda$) $\lambda_m$, so that
\begin{equation} \label{HGeigenvalues}
( H_A G_A) u^{(m)} = \lambda_m^2 u^{(m)}\, \ \ \ \  (G_A
H_A)v^{(m)}  =  \lambda_m^2 v^{(m)} \, ,
\end{equation}
and for simplicity introduce a normalization so that
\begin{equation}
(u^{(m)} \cdot v^{(m)})=1 \, .
\end{equation}
It is then possible to see that
\begin{equation}\label{crossrel}
 v^{(m)} =c_m G_A u^{(m)}\, \ \ \ \
u^{(m)}  =d_m H_A v^{(m)} \, .
\end{equation}
where the proportionality constants $c_m$, $d_m$ must satisfy the
condition
\begin{equation}
c_m d_m = \lambda_m^{-2}\, .
\end{equation}
A convenient choice to make is $c_m  = d_m = \lambda_m^{-1}$,
implying  that  $u^{(m)}$ and $v^{(m)}$ must be normalized so that
\begin{equation}
(v^{(m)}\cdot H_A v^{(m)}) = (u^{(m)}\cdot G_A u^{(m)}) =
\lambda_m \, .
\end{equation}
Next, we note that  since $G_A$ and $H_A$ are symmetric, $u^{(m)}$
and $v^{(m)}$  are respectively the right and left eigenvectors of
$H_A G_A$ respectively. This implies, together with our
normalization convention, the orthogonality condition
\begin{equation}\label{ortho}
(u^{(m)} \cdot v^{(n)}) =  \delta_{m n} \, \ \ \  \, ,
\end{equation}
as well as the spectral decomposition
\begin{equation}\label{specdecomp}
H_A G_A =\sum_m \lambda_m^2 u^{(m)} v^{(m)}{}^T \, .
\end{equation}
Using \rref{crossrel} and \rref{ortho}, we finally arrive at the
condition that
\begin{equation}\label{orthomats}
(v^{(m)}\cdot H_A v^{(n)}) = (u^{(m)}\cdot G_A u^{(n)}) =
\lambda_m \delta_{m n} \, .
\end{equation}
It is now straightforward to set up the symplectic transformation
matrix $X$: Letting $\phi_i$ be a column vector with all entries
set to zero except the $i$-th one,  one verifies by direct
substitution that the matrices
\begin{equation}
X = \sum_m  \phi_m u^{(m)}{}^T \, , \ \ \ \ Y = \sum_m \phi_m
v^{(m)}{}^T \, ,
\end{equation}
indeed satisfy \rref{transGH}. Finally, we can now express the
relation between the local and collective phase-space modes as
\beq \tilde q^{(m)}_{A} = \sum_{i \in A} u^{(m)}_i q_{i} \, , \ \
\ \ \tilde p^{(m)}_{A} = \sum_{i \in A} v^{(m)}_i p_{i} \, . \eeq

\subsection{Mode Participation Function}

Generically, the transformation connecting the initial modes to
the Williamson normal modes is not an orthogonal transformation.
This means that in general the mode functions $u^{(m)}$ and
$v^{(m)}$ for the $q$'s and the $p$'s may be very different in
shape. (For instance, it could be the case that the new $q$'s may
be fairly ``localized", with significant amplitude contributions
from only a small number of the old $q$'s, while the new $p$'s may
show more a ``collective" shape with more or less equal
contributions from the old $p$'s). Thus, it becomes convenient  to
define a function taking into account, on an equal footing, the
contribution from both the old $q$'s and $q$'s in  a given
Williamson mode.

To this end, we note the expansion of a given Williamson mode
creation operator in terms of local mode creation and annihilation
operators. If the local site creation operator is defined
as
\begin{equation}
a_i = \frac{1}{2}\left[ \xi_i q_i + \xi_i^{-1} p_i\right]
\end{equation}
where $\xi$ is some arbitrary dimensional parameter, the creation
operator for a given Williamson mode $m$ on one side can be
expanded as
\begin{equation}
a_{m} = \frac{1}{2} \sum_{i} (\xi_i^{-1} u^{(m)}_i + \xi_i
v^{(m)}_i)a_i + (\xi_i^{-1} u^{(m)}_i - \xi_i
v^{(m)}_i)a_i^\dagger \, ,
\end{equation}
In turn, this expression can be recast in the form
\begin{equation}
a_{m} =  \sum_{i}\sqrt{\mathcal{P}^{(m)}(i)} \left[ \cosh
\tau^{(m)}_i a_i + \sinh \tau^{(m)}_i a_i^\dagger \right] \, ,
\end{equation}
where
\begin{equation}
\mathcal{P}^{(m)}(i) \equiv u^{(m)}_i  v^{(m)}_i \, .
\end{equation}
and
\begin{equation}
\tau^{(m)}(i) \equiv \tanh^{-1}\left( \frac{ u^{(m)}_i - \xi_i^2
v^{(m)}_i}{u^{(m)}_i + \xi_i^2 v^{(m)}_i} \right)\, ,
\end{equation}
We also note from the definition of the modes that  $(u^{(m)}
\cdot v^{(m)})=1 $ so that $\sum_i \mathcal{P}^{(m)}(i) = 1$. From
the expansion of the mode creation operator, we see that the
function $\mathcal{P}^{(m)}(i) $ captures the weight of the of the
local site participation, invariant under local rescalings and
phase space rotations. We term it the  \emph{mode participation
function}

\subsection{Mode Mapping}

We now wish to relate the  the modes on the ``A"-side with the
modes on the ``B" side corresponding to the same symplectic
eigenvalue $\lambda_m >  \frac{1}{2}$. This can be done by noting
that from the isotropic condition \rref{isotropic},  the mode
structure can not only be obtained from the local covariance
matrices $M_A$ or $M_B$, but also from the correlation matrix $K$.
This in fact  proves to be advantageous if the correlations
between modes $A$ and $B$ are significant only for a small number
of modes, as in the case of short range interactions. Assuming
again no $q/p$ correlations, the correlation matrix between $A$
and $B$ can be expressed in the block diagonal form
\begin{equation}\label{formk}
K =\mathrm{Re}\ave{\eta_A\, \eta_B^T}=\left(%
\begin{array}{cc}
G_{AB} & 0 \\
0 & H_{AB} \\
\end{array}%
\right)\, , \
\end{equation}
where $G_{AB} = \ave{ q_A q_B^T }$ and $H_{AB} = \ave{ p_A p_B^T
}$.  Now, from \rref{isotropic} one obtains the relations between
$M_A$, $M_B$ and $K$
\begin{equation}
(J_A M_A)^2 + (J_A K)(J_B K^T) = \frac{1}{4}\openone \, ,
\end{equation}
\begin{equation}
(J_B M_B)^2 + (J_B K^T)(J_A K) = \frac{1}{4}\openone \, ,
\end{equation}
where $J_A$ and $J_B$ are the symplectic matrices in the $A$ and
$B$ sectors respectively. Substituting in the forms \rref{formma}
and \rref{formk}, we obtain the relations
\begin{mathletters}
\begin{eqnarray}\label{Ks}
H_A G_A & = & \frac{1}{4} - H_{AB} G_{AB}^T \, , \label{KA}\\
H_B G_B & = & \frac{1}{4} - H_{AB}^T G_{AB} \label{KB}\, .
\end{eqnarray}
\end{mathletters}

\begin{figure}
   \epsfxsize=3.0truein \epsfysize=2.4truein
\centerline{\epsffile{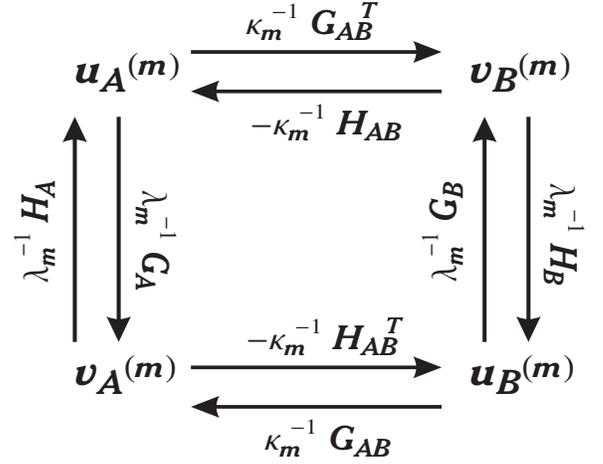}}
\medskip
\caption[\, ]{ Mode mapping.} \label{modemap}
 \end{figure}

Next, defining $\kappa_{m}^2$ to be the non-zero eigenvalues of
$-H_{AB} G_{AB}^T$, we thus see that the local symplectic
eigenvalues can also be expressed as
\begin{equation}
\lambda_m^2 = \frac{1}{4} + \kappa_m^2\, .
\end{equation}
Now, label the mode functions corresponding to the symplectic
eigenvalue $\lambda_m > \frac{1}{2}$ as $u_A^{(m)}$, $v_A^{(m)}$
and $u_B^{(m)}$, $v_B^{(m)}$ for sides $A$ and $B$ respectively.
According to \rref{Ks} and \rref{HGeigenvalues} we see that they
are solutions to the eigenvalue equations
\begin{mathletters}
\begin{eqnarray}\label{eveqa}
 H_{AB} G_{AB}^T u_A^{(m)} = - \kappa_m^2
u_A^{(m)}\, \label{eveqau}\\
 G_{AB} H_{AB}^T v_A^{(m)} = - \kappa_m^2
v_A^{(m)} \label{eveqav}\, .
\end{eqnarray}
\end{mathletters}
and
\begin{mathletters}
\begin{eqnarray}\label{eveqb}
 H_{AB}^T G_{AB} u_B^{(m)} = - \kappa_m^2
u_B^{(m)}\, \label{eveqbu}\\
 G_{AB}^T H_{AB} v_B^{(m)} = - \kappa_m^2
v_B^{(m)} \label{eveqbv}\, .
\end{eqnarray}
\end{mathletters}
Now, multiply both sides of say \rref{eveqau} on the left by
$G_{AB}^T$, to find from \rref{eveqbv} that $G_{AB}^T  H_{AB}
(G_{AB}^T u_A^{(m)})= - \kappa_m^2 (G_{AB}^T u_A^{(m)})$, thus
showing that $v_B^{(m)} \propto G_{AB}^T u_A^{(m)}$. A similar
procedure applied to all the above equations shows that
\begin{eqnarray}
v_B^{(m)} \propto G_{AB}^T u_A^{(m)}\ \  &\ \
v_A^{(m)} \propto G_{AB} u_B^{(m)} \\
u_B^{(m)} \propto H_{AB}^T v_A^{(m)}\ \ &\ \ u_A^{(m)} \propto
H_{AB} v_B^{(m)}\, .
\end{eqnarray}
The choice of the proportionality factors involved here is
constrained by the normalization conditions imposed on the mode
functions on both sides. A consistent choice preserving the
normalization on both sides is
\begin{eqnarray}
v_B^{(m)} =\frac{1}{\kappa_m} G_{AB}^T u_A^{(m)}\ \  &\ \
v_A^{(m)} =\frac{1}{\kappa_m}
G_{AB} u_B^{(m)} \\
u_B^{(m)} =-\frac{1}{\kappa_m} H_{AB}^T v_A^{(m)}\ \ &\ \
u_A^{(m)} =-\frac{1}{\kappa_m} H_{AB} v_B^{(m)}\, .
\end{eqnarray}

%
%
%

The above relations, together with the relations \rref{crossrel},
yield a systematic procedure by which all the mode functions on
both sides, corresponding to a given symplectic eigenvalue can be
constructed once a single mode function is found. This proves
particularly advantageous if the number of modes on one side is
considerably smaller that in the other. This construction is
summarized in figure \ref{modemap}.

\section{Harmonic Chain}

A linear harmonic chain of $N$ local oscillators laid out on a
circular topology may be modelled by canonical variables $(q_i,
p_i)$ with ${i= 1,..,N}$, with the dynamics given by a Hamiltonian
of the form
\begin{equation}\label{hamchain1}
    H = {E_0\over 2}\sum_{i=1}^{N} \left[ p_i^2+ q_i^2 -
\alpha\
    q_ni q_{i+\!1} \right]\, ,
\end{equation}
where we identify $q_{N+1} \equiv q_{1}$ and $q_{0} = q_{N}$ and
the dimensionless parameter $\alpha$ characterizing the strength
of the coupling between adjacent neighbor sites is strictly less
than unity. Note that such a Hamiltonian can be obtained from the
standard Hamiltonian of a chain with spring-like  nearest neighbor
harmonic couplings
\begin{equation}\label{hamchain2}
    H = \frac{1}{2}\sum_{i=1}^{N} \left[ \frac{\pi_i^2}{M}   + M
\omega^2 \xi_i^2 + K
    (\xi_ni -\xi_{i-1})^2 \right] \, .
\end{equation}
by means of the canonical variable re-scaling
\begin{equation}\label{cantransf1}
q_i = \sqrt{ M\omega \sqrt{1 + \frac{2K}{M \omega^2} }  } \, \
\xi_i \, ,
    \ \ \ \
 p_i = \frac{\pi_i}{\sqrt {M\omega \sqrt {1 + \frac{2K}{M \omega^2} }  }} \,
\end{equation}
and the identification
\begin{eqnarray}\label{paramtransf1}
    E_o & = & {\omega \sqrt{1 + \frac{2K}{M \omega^2} }} \, ,  \\
    \alpha  & = &
     \frac{ \frac{2K}{M\omega^2}} {1+ \frac{2K}{M\omega^2} }\, .
\end{eqnarray}
The last relation provides a restriction, $0<\alpha<1$, to the
possible values of the coupling constant in \rref{hamchain1}. The
limit of strong coupling between neighboring oscillators, $\frac
{2K}{M\omega^2} \to \infty$,  corresponds to $\alpha\to 1$, and
the weak coupling limit to $\alpha \to 0$.

The Hamiltonian \rref{hamchain1} can be brought to a normal form
by introducing a set of annihilation (creation) operators
$a(\theta_k)$ ( $a^\dag(\theta_k)$ )\,  satisfying the commutation
relations
\begin{equation}
[a(\theta_k), a^\dag(\theta_l)] = \delta_{kl} \, ,
\end{equation}
with the indexing angular variable $\theta_k$ playing the role of
a dimensionless wave number or pseudo-momentum and  taking the
values
\begin{equation}
\theta_k  =\frac{2 \pi k }{N}\, , \ \ \ \ (k =0,1,\ldots,N-1) \, .
\end{equation}
Defining the dispersion relation (in units of $E_o$)
\begin{equation}
\nu(\theta_k) \equiv \sqrt{ 1 - \alpha \cos \theta_k }\, ,
\label{disprel}
\end{equation}
and  expressing $q_n$ and $p_n$ in the form
\begin{eqnarray}
q_n & = & \frac{1}{\sqrt{ N}}\sum_k\,
\frac{1}{\sqrt{2\nu(\theta_k)}} \left[a(\theta_k) e^{i \theta_k
n}\! +\!\mathrm{h.c.}
\right]  \, ,\\
p_n & = & \frac{-i}{\sqrt{ N}}\sum_k
\sqrt{\frac{\nu(\theta_k)}{2}} \left[a(\theta_k) e^{i \theta_k
n}\! -\!
 \mathrm{h.c.}\right] \, ,
\end{eqnarray}
 the  Hamiltonian \rref{hamchain1} achieves the normal form
\begin{equation}\label{haminmodes}
    H = E_o \sum_k \nu(\theta_k)
    \left[a^\dag(\theta_k)a(\theta_k)+ \frac{1}{2} \right] \, ,
\end{equation}
which is  then diagonalized by the Fock states of the
creation/annihilation operators.

In particular, we will be interested in the ground state
$\ket{0}$, satisfying
\begin{equation}
a(\theta_k)\ket{0} = 0 \, ,
\end{equation}
for all $\theta_k$. For this  state,  the wave functions in the
coordinate and momentum representations assume the Gaussian form
\begin{eqnarray}
\psi_o(q) & \propto &     \exp\left[-\frac{1}{4}q^T
\covQ^{-1}q \right] \, , \\
\psi_o(p) & \propto &     \exp\left[-\frac{1}{4}p^T \covP^{-1}p
\right] \, ,
\end{eqnarray}
where the covariance matrices $G$ and $H$,  for $Q$ and $P$
respectively, satisfy the generalized uncertainty relation
$\covQ\covP = \frac{\openone}{4}$, with the entries defined by the
two-point vacuum correlation functions:
\begin{eqnarray}
\covQ_{i j} & =   \qave{0}{q_i q_j}  \equiv & \corQ{(i\!-\!j)} \, ,\label{vacorrs1}\\
\covP_{i j} & =   \qave{0}{p_i p_j} \equiv  & \corP{(i\!-\!j)} \,
. \label{vacorrs2}
\end{eqnarray}
Furthermore, since the state is Gaussian, higher moments of the
oscillator coordinates or momenta are expressible in terms of the
two-point correlation functions. Thus, the relevant physical
information associated with the vacuum is contained in the
correlation functions $\corQ{(i\!-\!j)}$ and $\corP{(i\!-\!j)}$,
which we now study.

\subsection{ Vacuum Correlation Functions and Three Regimes of Behavior}
\label{corrbehav}

The general entanglement behavior of the partitioned harmonic
chain is dictated by the behavior of the  correlation functions
$g_l$ and $h_l$ defined in Eqs. \rref{vacorrs1} and \rref
{vacorrs2}. Their explicit form is given in terms of the
dispersion relation \rref{disprel} by
\begin{eqnarray}\label{corresums}
\corQ{l}^{(N)} & \equiv & \frac{1}{2 N
}\sum_k\frac{1}{\nu(\theta_k)}\cos (l\theta_k)  \, ,\\
\corP{l}^{(N)}  & \equiv & \frac{1}{2 N }\sum_k \nu(\theta_k)\cos
(l\theta_k) \, ,
\end{eqnarray}
and consistent with the translational invariance of  the
Hamiltonian, we note their dependence only on the separation
$l=(i-j) \mod N$. In the limit $N \rightarrow \infty$ with
$\alpha$ fixed, these expressions  yield the Riemann sum for the
integral of the argument as a function of a continuous $\theta$
ranging from $0$ to $2\pi$,  with the replacement of the factor of
$N^{-1}$ in front by $(2 \pi)^{-1}$. The correspondence with the
continuum one-dimensional scalar field theory can also be obtained
from these expressions by taking appropriate limits, as discussed
in the appendix.

Now, for a fixed value of the strength parameter $\alpha$ and for
sufficiently large values of the chain size $N$, the behavior of
these functions in terms of $l$  becomes independent of $N$ and
reproduces the $N \rightarrow \infty$ behavior, which can be
expressed exactly in terms of hypergeometric functions as
\begin{eqnarray}
\corQ{l}^{(\infty)} & =& \frac{z^l}{2 {\mu} }\,
\left(\!\begin{array}{c} l\! -\!\frac{1}{2} \\ l \end{array}\!
\right) {}_2\! F_{1}\left({\frac{1}{2}},l\!+\!{\frac{1}{2}}
,l\!+\!1,{ z^2}\!\right) \label{hyperg}
 \\
\corP{l}^{(\infty)} & =& \frac{{\mu} z^l}{2}\, \left(\!
\begin{array}{c} l\! -\!\frac{3}{2} \\ l \end{array}\!
\right){}_2\! F_{1}\left(-\!{\frac{1}{2}},l\!-\!{\frac{1}{2}}
,l\!+\!1,{ z^2}\!\right) \label{hyperh}
\end{eqnarray}
where $\left(\begin{array}{c} a \\ b \end{array}\right)$ are
binomial coefficients expressed in terms of Gamma functions and
$z$ and $\mu$ are given by
\begin{equation}\label{defz}
z \equiv \frac{1 - \sqrt{1 -\alpha^2}}{\alpha} \, , \ \ \ \ \ \mu=
\frac{1}{\sqrt{1 + z^2}}\, .
\end{equation}
This behavior in the large $N$ limit will serve as a reference in
analyzing the behavior for finite $N$.

\begin{figure}
   \epsfxsize=3.4truein
\centerline{\epsffile{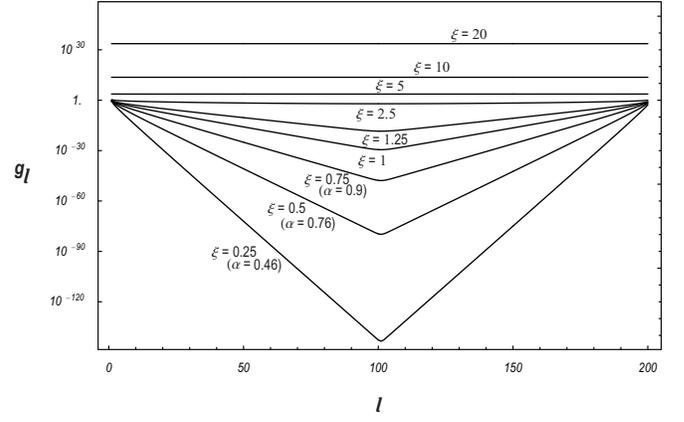}}
\medskip
\caption[\, ]{The vacuum $g_l = \langle q_o q_l \rangle$
correlation function as a function of $l$ for different strength
parameters where $\alpha=\tanh{2\xi}$. }\label{fig:gl}
 \end{figure}

\begin{figure}
   \epsfxsize=3.4truein
\centerline{\epsffile{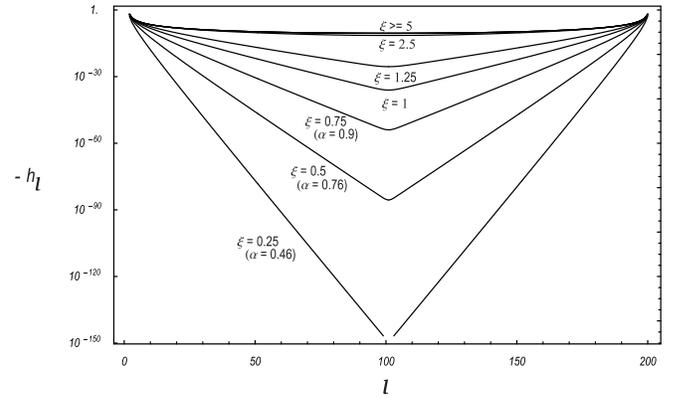}}
\medskip
\caption[\, ]{ The negative of the vacuum correlation function
$h_l = \langle p_o p_l \rangle$
 as a function of $l$ ($l \geq 1$) for different
strength parameters.}\label{fig:hl}
 \end{figure}

\begin{figure}
   \epsfxsize=3.0truein
\centerline{\epsffile{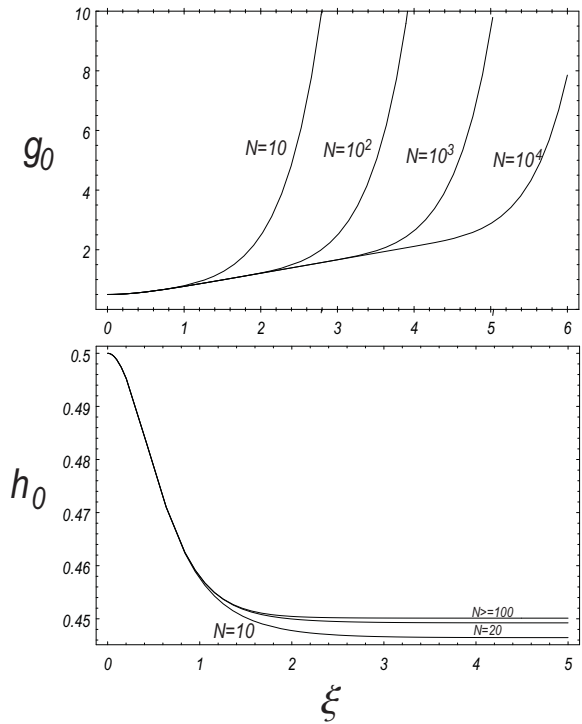}}
\medskip
\caption[\, ]{Variances in $q$ and $p$ as a function of coupling
strength for different values of $N$. }\label{fig:g0h0}
 \end{figure}

The shape of the correlations $g_l$ and $h_l$ as a function of the
separation $l$ and the coupling strength is depicted in figures
\ref{fig:gl} and \ref{fig:hl} for a fixed value of $N$. Similarly,
we have plotted in \ref{fig:g0h0} the behavior of their values at
$l=0$ as a function of the coupling strength for different values
of $N$. For these plots, we have found it convenient to introduce
an ancillary hyperbolic angle $\xi$, implicitly defined by its
relation to the variables $\alpha$ and $z$:
\begin{equation}
z = \tanh \xi \, \ \ \ \ \alpha = \tanh 2 \xi
\end{equation}
This becomes a convenient parameter as for small values of
$\alpha$, the appropriate expansion parameter is $z$ itself, in
which case $z \simeq \xi$ for $\xi \ll 1$; similarly, as $\alpha$
approaches unity, $\xi$ provides a logarithmic scale for this
approach, with $\xi \simeq \frac{1}{4}\ln(1 -\alpha)/2 $.

Let us first analyze the case $N= \infty$. For  weak coupling,
$\alpha \sim z \ll 1 $, the hypergeometric functions behave as $ 1
+ o(z^2)$; thus, with the Stirling approximation for the binomial,
we obtain the weak coupling behavior
\begin{equation}\label{correlength}
g_l \simeq \frac{1}{2\sqrt{\pi}}\, l^{-\frac{1}{2}}z^l \, , \ \ \
\ \ h_l  \simeq -\frac{1}{4\sqrt{\pi}}\,  l^{-\frac{3}{2}}z^l \, ,
\end{equation}
for $l \gg 1$. The weak coupling correlation functions are
therefore of short-ranged, exponential behavior, and characterized
by the correlation length
\begin{equation}
l_c = \frac{1}{-\ln(z)} \, .
\end{equation}
On the other hand, in the limit $\alpha \simeq z \rightarrow 1$
with fixed $l$, the $g^{(\infty)}_l$ correlation function
diverges, and is determined by the asymptotic behavior of the
hypergeometric function
\begin{eqnarray}\label{gasymp}
&{}& {}_2F_{1}\left({\frac{1}{2}},l\!+\!{\frac{1}{2}} ,l\!+\!1,{
z^2}\!\right) \rightarrow -\frac{\Gamma(l+1)}{\Gamma(1/2)\Gamma(l
+ 1/2)}\\ \nonumber &{}& \times
\biggl[\ln\left(\frac{1-z^2}{4}\right)  +
\psi\left(l\!+\!{\frac{1}{2}}\right) + \gamma\biggr]( 1 + O(1-z))
\, .
\end{eqnarray}
where $\gamma$ is the Euler-Mascheroni constant and $\psi(n)$ is
the DiGamma function. With the asymptotic expansion $\psi(l+1/2)
\simeq \ln l  $ for large $l_c \gg l \gg 1$, this yields for $z
\simeq 1$,
\begin{equation}\label{hasymp}
g_l \simeq -\frac{1}{\sqrt{2} \pi} \ln\left(\frac{1\!-\!z}{2}\, l
\right) \, ;
\end{equation}
in the case of the correlation function $\corP{l}$, one finds the
$\alpha$ independent limit
\begin{equation}
\lim_{\alpha \rightarrow 1} \corP{l}  = -\frac{\sqrt{2}}{\pi}
\frac{1}{ 4 l^2 -1 } \, .
\end{equation}

Next we consider  finite-size effects. The first consideration has
to do with the range  the above limiting expressions given a
finite size of the chain.  In this respect, the exponentially
decaying behavior is not expected to hold once the correlation
length becomes of the order of half the chain size; similarly, the
strong coupling expansion for $g_l$ is valid whenever the function
is positive.  For sufficiently large $N$, one should expect these
conditions to be met at values of $\alpha$ which are close to
unity, in which case $1-z \simeq  \sqrt{2(1 -\alpha)}$. Then the
conditions $l_c < N/2$ and $ \ln\left((1-z)N /4\right) < 0 $, lead
essentially to the same transitional condition between short-range
and long range correlation; the transition from short -ranged to
long-ranged behavior then happens when  a transitional correlation
size scale, which we define as
 \begin{equation}
N_t(\alpha) = \sqrt{\frac{2}{1 -\alpha}}  \, ,
\end{equation}
becomes of the order of $N$. Note incidentally that for $N_t \gg
1$, the plotting parameter $\xi$ is related to $N_t$ according to
\begin{equation}
\xi \simeq \frac{1}{2} \ln N_t \, .
\end{equation}

The second consideration has to do with the magnitude of the
correlations for $\alpha$ close to unity,  which is set by the the
diverging $qq$ correlation function and specifically its behavior
at $l=0$. For finite but large $N$, the value of $g_l$ can be
approximated by the $N=\infty$ expression plus corrections in
powers of $N^{-1}$ arising from the error between  the sum
\rref{corresums} and the corresponding Riemann integral obtained
when $N \rightarrow \infty$.  The relevant correction comes from
the contribution of the $\theta_k = 0$ mode in \rref{corresums},
which is not accounted for in the  Riemann integral when the limit
$\alpha \rightarrow 1$ is taken first. This contribution yields an
$O(N^{-1})$ additional term, so that for $\alpha \rightarrow 1$,
the finite size $g_l$ is  approximated by
\begin{equation}
\corQ{l}^{(N)} \simeq \corQ{l}^{(\infty)} + \frac{1}{2 N \sqrt{1 -
\alpha}} \, .
\end{equation}
Since the  correction arises from the contribution to the sum from
the $\theta_k = 0$ eigenmode,  it accounts for a strong collective
effect due the finite size of the chain. This correction dominates
the coupling strength behavior of the $g_l$ correlation functions
above a critical value of $\alpha$, or, for fixed $\alpha$, below
a critical value of $N$. This critical value is determined by
requiring that the correction term should become of the magnitude
of the logarithmic term depending on the coupling strength in the
strong coupling $N=\infty$ correlation function, i.e. ,
\begin{equation}
\frac{1}{2 N \sqrt{1 - \alpha}} =
-\frac{1}{\sqrt{2}\pi}\ln\left(\frac{1 - z(\alpha)}{2} \right)
\end{equation}
Taking $\alpha$ close to unity yields finally a critical chain
size value  $N_c$
\begin{equation}
N_c(\alpha) \sim \frac{N_t(\alpha)}{ \ln N_t(\alpha)} \, .
\end{equation}

The above considerations allow us to distinguish three regimes of
behavior for a chain size of a large and fixed value of $N$, for
convenience to be referred to as the type $I$, $II$ or  $III$
regimes:  the type $I$, weak coupling regime, is determined by the
condition $N \gg N_t(\alpha) > N_c(\alpha)$, and is characterized
by short-ranged correlations; when the condition $N_t(\alpha) \sim
N$ is met, we enter the intermediate, type $II$ long-range regime,
in which the scale of correlations is of order and $\ln
N_t(\alpha)$ and the correlations decay  logarithmically as a
function of $l$; finally,  the condition $N_c(\alpha) > N$
determines the long-range, type $III$ regime, with the same
logarithmic decay as a function of the distance but in which the
scale of the correlations behaves as $N_t(\alpha)/N$. The
distinction between these three regimes will serve a guideline for
the characterization of the different regimes of entanglement as
well.

\section{An Illustrative Exactly Soluble Case}
\label{singlemode}

Before proceeding with the numerical results in our paper, it may
be useful to consider a simple example that is easy to solve and
shows very general qualitative features of the dominant
entanglement structure between two complementary regions of a
chain. In this case we wish to understand the entanglement of one
oscillator of the chain versus the rest of the chain. For such a
partition, it is quite easy to obtain the degree of entanglement
by looking at the local covariance matrix of the single mode,
which is given by
\begin{equation}
M_A = \left(%
\begin{array}{cc}
  g_0 & 0 \\
  0 & h_0 \\
\end{array}%
\right)\, .
\end{equation}
The single symplectic eigenvalue of the matrix is $\lambda =
\sqrt{g_0 h_0}$ and thus the degree of entanglement between the
oscillator and the remainder of the chain is given is given by
$S(\sqrt{g_0 h_0})$ where $S$ is defined by Eq. \rref{entropy}.

To understand the behavior of the degree of entanglement we look
at the relevant approximations for the three regimes discussed in
section \rref{corrbehav}. First, expanding the hypergeometric
functions in \rref{hyperg} and \rref{hyperh} in a power series in
$z$, we obtain for the symplectic eigenvalue the weak, type $I$
regime,
\begin{equation}
\lambda_I = \frac{1}{2} + \frac{z^2}{8} + O(z^4) \, .
\end{equation}
Next, for strong coupling, $N_t(\alpha) > N$, and assuming $N \gg
1$, we  approximate $h_0$ and $g_0$ by their limiting behaviors
for $\alpha \rightarrow 0$. For  $h_0$ the limiting value is
$\sqrt{2}/\pi$, while for $g_0$ we can use the asymptotic
\rref{gasymp}  to show that
\begin{equation}
\corQ{0}(\alpha) \simeq \frac{1}{ \sqrt{2} \pi}\ln\left( 4
N_t(\alpha)\right) \, ,
\end{equation}
where $N_t(\alpha) = \sqrt{2 /(1-\alpha) }$ as defined previously.
Thus the  symplectic eigenvalue becomes
\begin{equation}
\lambda_{II} \simeq \frac{1}{\pi}\sqrt{\ln(4 N_t(\alpha) )} \, ;
\end{equation}
this accounts for the type $II$, long-range regime with a
logarithmic scale of the correlations. Finally, if, we take
$\alpha$  large enough so that $N_c(\alpha) \gg  N$, then the
$g_0$ correlation function can be approximated by the single
contribution of the collective $\theta_k = 0$  of the whole chain
\begin{equation}
\corQ{0}(\alpha) \simeq \frac{1}{2 N \sqrt{1 - \alpha}} =
\frac{1}{2  \sqrt{2}}\left(\frac{N_t(\alpha)}{N}\right)\, .
\end{equation}
This yields the symplectic eigenvalue
\begin{equation}
\lambda_{III} \simeq \sqrt{\frac{N_t(\alpha)}{2 \pi N}} \,
\end{equation}
characterizing the  strong long-range (type $III$) regime.

\begin{figure}
   \epsfxsize=3.4truein
\centerline{\epsffile{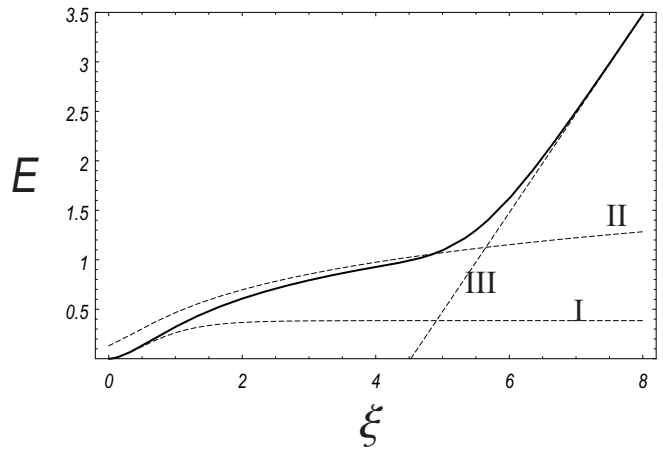}}
\medskip
\caption[\, ]{Entanglement of a single oscillator versus the rest
of the chain  as a function of coupling strength for a chain size
$N=10^4$ (solid). The dashed lines show the approximations of Eq.
\rref{singlentapps}: weak (I), strong transitional (II), and
strong (III).}\label{fig:singlent}
 \end{figure}

With these approximations we use the following expansions for the
von-Neumann entropy \rref{entropy} as a function of $\lambda$
\begin{equation}
S(\lambda) \simeq \left\{ \begin{array}{cc} (\lambda -\frac{1}{2})\left[ 1 - \ln (\lambda -\frac{1}{2}) \right] \, , &  \ \ \lambda -\frac{1}{2} \ll 1 \\
{} & {} \\
1 + \ln \lambda \, ,   & \ \ \lambda  \gg 1 \end{array} \right. \,
.
\end{equation}
This yields, for the characteristic behavior of the degree of
entanglement in the three regimes, the expressions
\begin{equation}\label{singlentapps}
E \simeq \left\{ \begin{array}{cc} \frac{z^2}{8}\left[ 1 - \ln \frac{z^2}{8} \right] \, ,  &  \ \ I\ :\ N_t(\alpha) \ll N \\
{} & {} \\
1 + \frac{1}{2}\ln \frac{\ln 4 N_t(\alpha)}{\pi^2}\ \, ,  &  \ \  II\ :\ N_t(\alpha) \sim N \gtrsim N_c \\
{} & {} \\
1 + \frac{1}{2}\ln  \frac{ N_t(\alpha)}{2 \pi N}\ \, ,  &  \ \ II\
:\ N_c(\alpha) \gg N\ \ \  \, .
\end{array} \right.
\end{equation}
These three approximations  compared quite with numerical
computations of the entropy as a function of coupling strength in
Fig \ref{fig:singlent} for a chain size of $N = 10^4$. We note
that  the only importance dependence on the chain size for fixed
coupling strength comes in the very strong regime, in which case
the entanglement decreases logarithmically with $N$.

Correlated with the three regimes of behavior are the shapes of
the $u$ and $v$ mode functions for the entangled mode on the
complement of the chain. These are easily  expressed in terms of
the correlation functions $g_l$, $h_l$ using the mode-mapping
diagram Fig. \ref{modemap}; The $l$-th components are simply given
by
\begin{eqnarray}
&v_l = \frac{g_l}{\sqrt{g_o h_o -\frac{1}{4}}} & \, \ \ \ \ \,
\\
&u_l = -\frac{h_l}{\sqrt{g_o h_o -\frac{1}{4}}} & \, \ \ \ \
l=1,..,N-1 \, .
\end{eqnarray}
Thus, the shapes of $u$ and $v$ are easily inferred from figures
\ref{fig:gl} and \ref{fig:hl}: in the weak regime, the mode
functions  exhibit rapid exponential decay away from the two sites
adjacent to the singled-out oscillator; in the long-range
transitional regime, the $u_l$ mode, being proportional to $h_l$,
decays as the inverse squared of the distance from the edge, while
the $v_l$ mode decays logarithmically; a similar behavior is
exhibited in the strongest long-range regime, except that the
$v_l$ mode except that the logarithmic behavior is essentially
suppressed. Thus we see that in the strongest regime the entangled
mode at the complement shows a collective correlated behavior in
the oscillator momenta and a power-law correlation in the
oscillator displacements. Two distinct limiting behaviors in
$\alpha$ may then be inferred for the mode participation function
of this mode, based on the asymptotics of the correlation
functions, namely
\begin{equation}
P_l \sim \left\{ \begin{array}{cc} l^{-2} z^{2l} \, , &  \ \ \alpha \rightarrow 0 \\
{} & {} \\
l^{-2} \, ,   & \ \ \alpha \rightarrow 1 \end{array} \right. \, .
\end{equation}

\section{Structure of ground state entanglement:
Qualitative study and numerical results}
 \label{qualitative}

\begin{figure}
\epsfxsize=3.2truein  \centerline{\epsffile{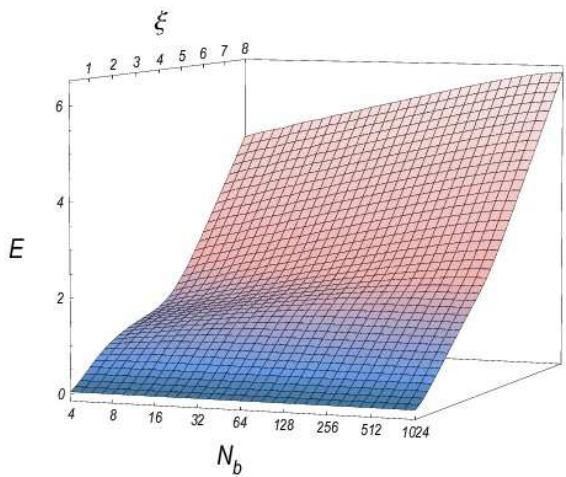}}
\medskip
\caption[\, ]{Total entanglement as a function of coupling
strength (in terms of $\xi$) and $N_b$ (log-2 scale) for a chain
of size $N = 2048$. } \label{fig:3dtotent}
\end{figure}

\begin{figure}
\epsfxsize=2.5truein  \centerline{\epsffile{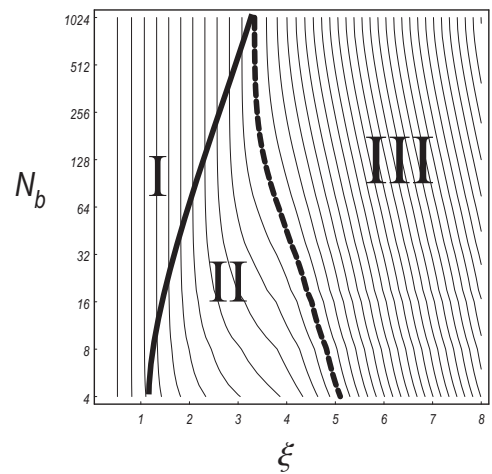}}
\medskip
\caption[\, ] {The three different qualitative regimes of
entanglement as inferred from level curves of total entanglement
in Fig. \ref{fig:3dtotent}. }\label{fig:entphases}
\end{figure}

We turn then to a qualitative study  of the entanglement behavior
of the harmonic chain vacuum and the underlying modewise
structure, based on numerical results.  We shall  be interested in
a system of $N$ oscillators, of which the first contiguous $N_b
\leq N/2$  constitute a subsystem that we shall refer to as the
``block", while  the remaining $N - N_b$ will constitute what we
shall refer to as the ``complementary block".  From the
translational symmetry of the problem, the starting point of the
block on the chain is completely arbitrary.

Our first survey has to do with the overall behavior of the degree
 of entanglement of the block and its complement as a function of
 the relevant properties: the size of the block $N_b$, the size of the chain $N$, and the
 coupling strength. This overall behavior has been plotted in
  Fig. \ref{fig:3dtotent}, for a chain of size $N = 2048$,
  sweeping the block size up to $N_b = N/2$ and the coupling
  strength so that $N_c > N$.
As was seen in the previous section, the behavior of the degree of
entanglement  as a function of the coupling bears the signature
 of the three regimes  outlined in section \rref{corrbehav}.
 This signature is also evident from Fig. \ref{fig:3dtotent}
 when we look at slices of constant $N_b$: As a function of
 $\xi= \tanh(2 \alpha)$, the entanglement rises from zero in
 the weak coupling regime  to the characteristic plateau of
 the type $II$ regime, which is more or less centered around
 the  value for which the transitional scale
  is of the order of the chain size
  (in our case around $\xi \simeq 3.2$); as $\xi$ is further
  increased, we see again the characteristic $\sim \ln N_t$
  (i.e., linear in $\xi$) behavior of the degree of entanglement
  for the type $III$ regime.
As we sweep then  $N_b$ and look at the entanglement as a
 function of  $\xi$, it then becomes possible to create an
 analog of a phase diagram on the $\xi-N_b$ plane for the
qualitative behavior of the degree of entanglement;  such a
diagram is show in Fig. \ref{fig:entphases} and was constructed on
the basis of the level curves of equal entanglement obtained to
Fig. \ref{fig:3dtotent}.

\begin{figure}
\epsfxsize=3.25truein  \centerline{\epsffile{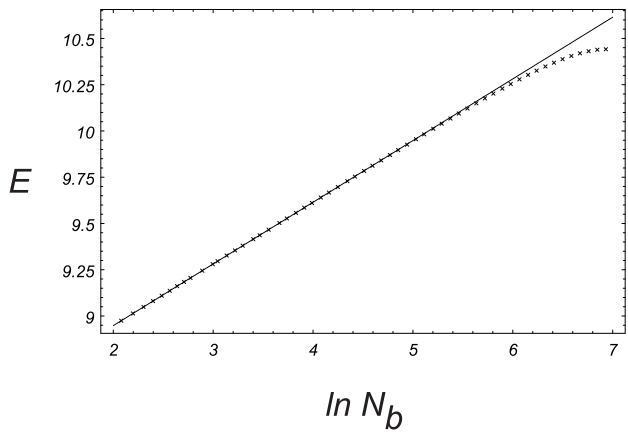}}
\medskip
\caption[\, ] {Entanglement as a function of $\ln N_b$ for
$N=2048$, $\xi =12$ fitted to a straight line of slope $1/3$.
}\label{fig:ln3}
\end{figure}

\begin{figure}
\epsfxsize=3.4truein  \centerline{\epsffile{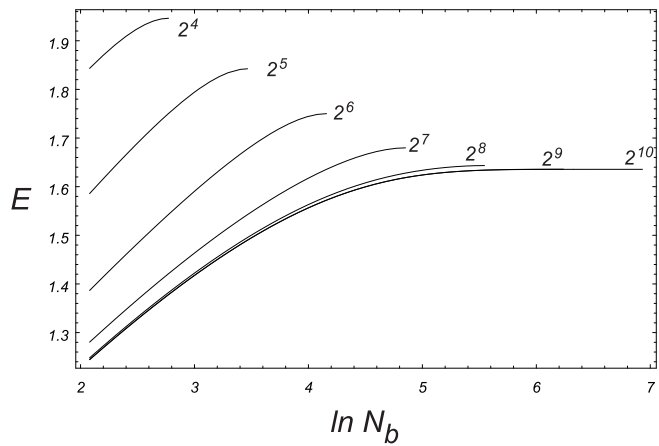}}
\medskip
\caption[\, ]{Entanglement as a function of block size for
different values of the chain size  $N$ ($\xi \simeq
3.5$).}\label{fig:entasnc}
\end{figure}

Continuing with  the  behavior of the total entanglement as a
function of the block size,  we again find
 distinct  characteristic behaviors depending on the  specific
  regime probed in the "phase diagram": for the type $I$ regime,
   the total degree of entanglement remains essentially constant
    as  $N_b$ is varied. Clearly, this is a manifestation of the
     short-ranged correlations characteristic of this regime,
     and suggests that  entanglement between the block and its
      complement in this regime is due essentially to edge effects.
       More interesting, however, is the behavior in the strong type-$II$
       and type-$III$ regimes, where long range correlations are present
       and where therefore one should expect significant contributions
       from the bulk of the block. Indeed, one finds that in both of
       these regimes the degree of entanglement shows a logarithmic
       dependence on the size of the block--the one-dimensional
       analogue of the black hole area theorem. As shown in
       Fig. \ref{fig:ln3}, for sufficiently
       large values of the coupling and $N$, and for $N_b \ll N$
       the  dependence takes the form
       \begin{equation}
       E \simeq E_0(\alpha,N) + \frac{1}{3} \ln N_b
       \end{equation}
reproducing the  predictions of conformal field theory for the
boson universality class \cite{vidalrev,holzhey}. It is also
interesting to note the behavior of the  value $E_0(\alpha,N)$ as
a function of the chain size $N$ for fixed $\alpha$, which shows a
similar behavior to that of the entanglement of the single
oscillator mentioned in the previous section. As shown in Fig.
\ref{fig:entasnc}, for large fixed value of $\alpha$, the $\propto
\ln N_b$ curves  decrease in height with $N$ until a critical
value of $N$; below this value the $\propto \ln N_b$ behavior
saturates at a value that is independent of the chain size.

\begin{figure}
\begin{center}
\includegraphics[width=9cm]{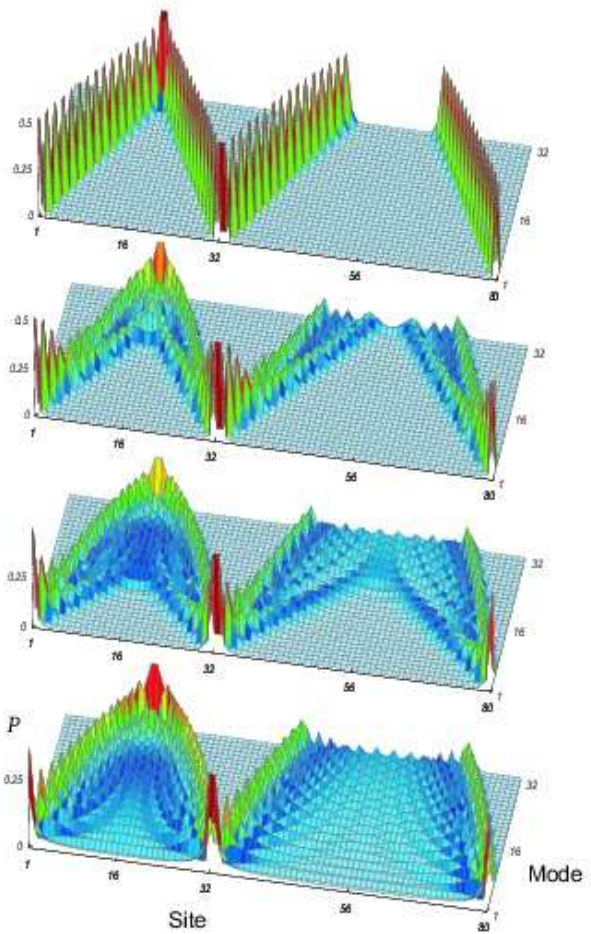}
\end{center}
\medskip
\caption[\, ]{Site participation function for the entangled
Williamson normal modes in a chain of $N=32+48$ contiguous
oscillators, at four values of the coupling strength parameter
(from top to bottom): $\alpha=0.1$, $0.6$, $0.9$, $1-10^{-6}$. The
mode shapes are ordered front to back according to the symplectic
eigenvalue of the respective mode, with the dominant mode at the
front. }\label{fig:allmodes}
\end{figure}

\begin{figure}
\begin{center}
%
%
\epsfxsize=3.5truein  \centerline{\epsffile{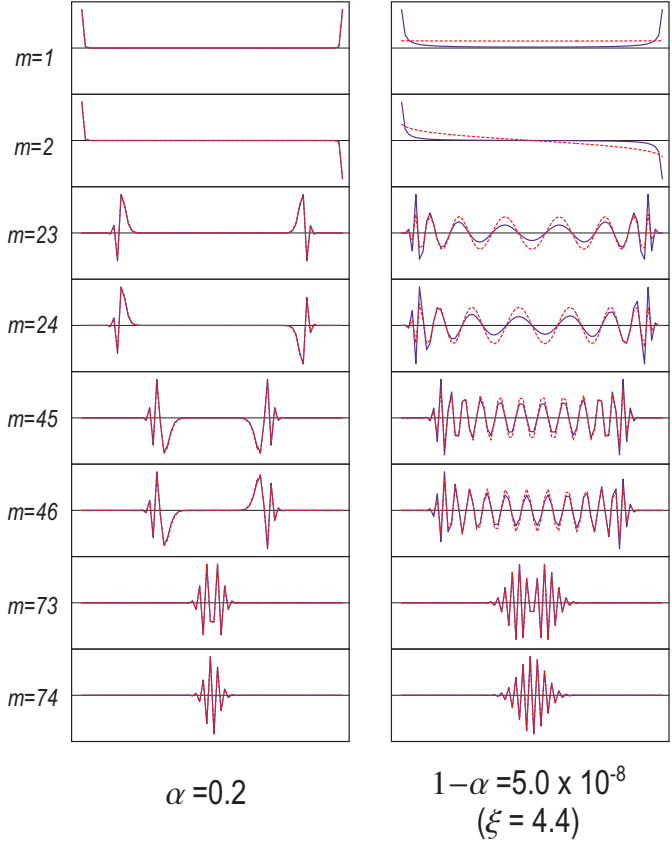}}
\end{center}
\caption[\, ]{ A sample of Williamson mode shapes (solid for
$u^{(m)}$, dashed for  $v^{(m)}$) for a chain of size $N_b = 74$
($N = 500$), and for weak and strong coupling. Mode ordering is
according to decreasing contribution to the total entanglement. }
\label{fig:uvmodes}
\end{figure}

We turn then to the question of ``where does the entanglement
comes from ?", from the point of view of the modewise entanglement
structure  of the harmonic chain vacuum. In conformity with the
different regimes of behavior entailed by the  correlation
functions $g_l$ and $h_l$ ,  different behaviors are reflected in
the shapes of the Williamson modes, as we illustrate in Fig.
\ref{fig:allmodes}, which shows the site participation function
for all the entangled Williamson normal modes of both the block
and its complement. Figure \ref{fig:uvmodes} also shows a small
sample of the mode shapes for the $u^{(m)}$ and $v^{(m)}$ modes,
for various values of $m$, and at two different coupling
strengths. A number of general features may then be identified
regarding the mode structure:

The first feature, which is evident from Fig, \ref{fig:uvmodes},
has to do with the parity of the mode functions $u^{(m)}$ and
$v^{(m)}$, due to a parity selection rule to be discussed in
section \rref{weakanal}. Indexing the modes as $m =1,...N_b$, in
increasing order of their contribution to the total entanglement,
we find that in all circumstances entangled modes of either the
chain or its complement have a definite parity $(-1)^{m+1}$, under
reflections with respect to the center of the respective block.
Thus,  for each entangled mode-pair, the modes on both sides of
the chain have the same parity, and  the mode-pair that
contributes the most amount of entanglement always involves even
mode functions on both sides of the chain. This feature shows that
while it may be argued that the entanglement between two regions
of the chain is primarily due to edge effects, it nevertheless
involves non-local behavior within each region.

\begin{figure}
\epsfxsize=3.0truein  \centerline{\epsffile{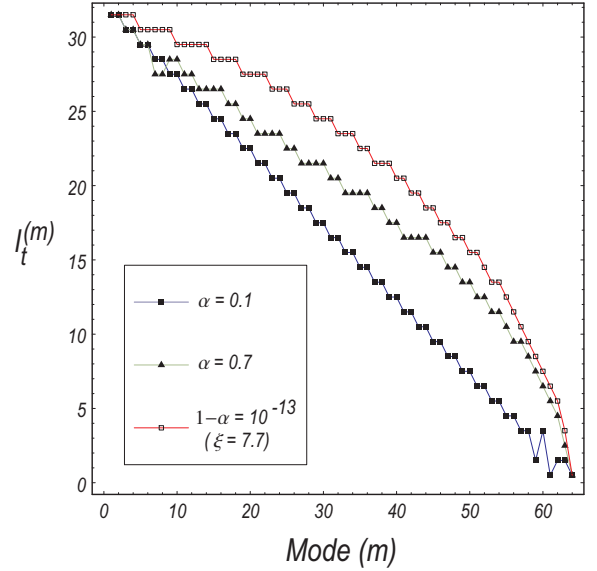}}
\medskip
\caption[\, ]{Turning point location as a function of mode number,
for three values of the coupling strength ($N_b = 64$, $N =
160$).}\label{fig:turningpoint}
\end{figure}

A second  feature of the mode structure that is present in all
circumstances is the existence of a characteristic turning point,
henceforth denoted by $l_t^{(m)}$, which we define as the location
(measured from the center of the block) of the oscillator for
which the participation function $\mathcal{P}_i^{(m)}$ is maximal
for the $m$-th mode. The   general behavior of this turning point
as a function of the mode number and coupling strength is shown in
Fig. \ref{fig:turningpoint}.   What is observed is that
$l_t^{(m)}$ is quite similar to the turning point of semiclassical
solution to a wave equation: generally,  the mode participation
shows an exponential-like decay beyond the turning point, so that
all the mode activity becomes effectively confined to site
distances up to $l_t^{(m)}$ from the block midpoint; furthermore,
insider this region, we  find a less dramatic decay of the mode
amplitudes from the turning point, achieving the minimum amplitude
at the midpoint. However, in contrast to semiclassical solutions,
what we find is an anti-correlation between the amplitude and the
local wavelength:  the longest wavelengths are present at the
midpoint, whereas the oscillations become of the order of the
lattice spacing at the turning point.

\begin{figure}
\begin{center}
\epsfxsize=2.5truein  \centerline{\epsffile{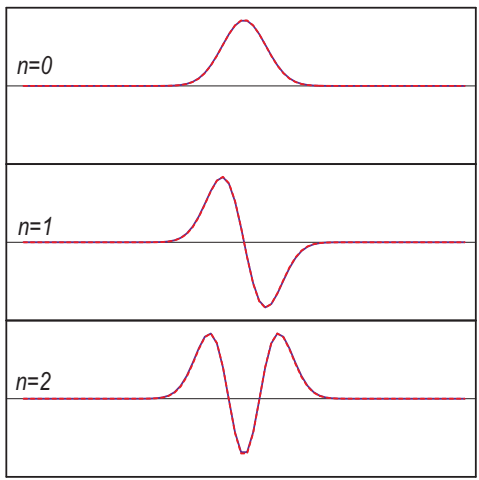}}
\end{center}
\caption[\, ]{ Mode shapes for the first three innermost modes for
the strong regime case depicted in Fig. \ref{fig:innersamples},
demodulated by the oscillating factor $(-1)^i$. Here $n = N_b -
m$. } \label{fig:innersamples}
\end{figure}

The last general feature that we shall discuss is observed more
prominently as the coupling strength increases, and has to do with
the typical wavelength of oscillation of the mode over the
majority of the interior region bounded by the turning points.
This characteristic wavelength decreases with the depth ($m$) from
around the order of of the chain size, for the outer mode, to  the
lattice spacing for the innermost modes. Interestingly, we also
find a ``dual" structure to the mode shape oscillations if the
$u^{(m)}_i$ and $v^{(m)}_i$ mode entries are multiplied by the
alternating factor $(-1)^i$ (corresponding to a shift of $\theta =
\pi$ in their discrete Fourier transform), and the modes are
ordered from the inside out. Using the index convention $n = N_b
-m$, we find that as the coupling strength increases, the mode
shapes for the lowest values of $n$ in this hierarchy of
``demodulated" modes become virtually indistinguishable from
harmonic oscillator wave functions of the  corresponding value of
$n$, with a variance of order $\sqrt{N_b}$ (Fig.
\ref{fig:innersamples}). More generally, for the whole hierarchy
the index $n$ labels the number of nodes in the demodulated mode
functions.

\begin{figure}
   \epsfxsize=3.4truein  \centerline{\epsffile{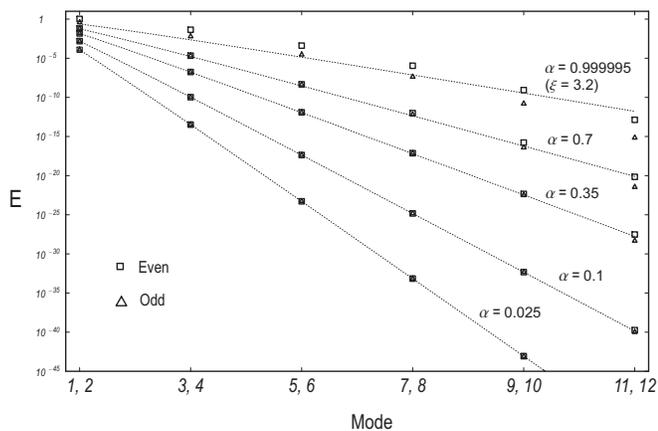}}
\medskip
\caption[\, ]{Entanglement as a function of mode number in the
weak regime, for different values of the coupling constant. Here
$N_b = 12, N=500$.}\label{fig:weakspecs}
\end{figure}

With these considerations in mind, we turn next to the
characterization of the regimes of behavior discussed earlier from
the standpoint of the mode structure.
Beginning with the weak regime, we find that the modes for both
the block and the complement fall into pairs of even and odd
combinations of oscillator sites that are precisely localized in
increasing distance from the edge of the block (Fig.
\ref{fig:turningpoint}), so that a  mode  involving a certain
oscillator pair of the  block is entangled with a  mode of the
complement involving the two oscillators' specular images  with
respect to reflections about the interface. This ``wedge-like"
structure  in the mode distribution of both sides of the chain
observed in Fig. \ref{fig:allmodes} can be understood in  the
$\alpha \rightarrow 0$ limit, where the basis of localized
oscillator sites serves an eigenbasis for the whole chain; then,
the leading  effect of any infinitesimal coupling between
contiguous oscillators is to recombine the localized modes so as
to produce a mode basis consistent with the parity selection
rules. We refer the reader to section \ref{weakanal} for more
careful analysis discussion in this respect.

As  we increase the coupling strength, and hence the correlation
length $l_c(\alpha)$, the mode shapes become distorted to the same
extent that the two participating sites lie within a distance from
each other that is less than or equal to $l_c(\alpha)$. Thus the
distortion of the modes  proceeds from the inside modes of the
block and then outwards as the coupling strength is increased.
This distortion involves a gradual diffusion of the mode
participation towards the interior region bounded by the modes, as
well as the establishment of a characteristic wavelength of
oscillation within this region. As the frequency spectrum of the
modes becomes sharper, the width of the interior region of the
mode becomes broadened, thus accounting for the gradual outward
curving of the turning point location seen in Fig.
\ref{fig:turningpoint}.

Together with this mode-shape distortion, we also find the
behavior of the entanglement contribution of the modes to be
dictated by  the correlation length in the weak regime.   In Fig.
\ref{fig:weakspecs} we have plotted on a logarithmic scale the
dependence of the entanglement on the mode number in the weak
regime. As stated earlier,  in this  regime the modes come in
pairs of definite-parity combinations of oscillator sites at a
precise distance from the edge; letting $d_m$ be this distance
($d_m = m/2$ for $m$ even, $(m+1)/2$ for $m$ odd) we have found
that the entanglement is to leading order  degenerate between
modes of opposite parity and the same  $d_m$, independent of $N_b$
(for $N_b >2$),  and of exponential fall-off   with $d_m$
according to
\begin{equation}\label{weakent}
E_{m}(\alpha) \simeq \left(\frac{z(\alpha)}{4} \right)^{2(2 d_m-
1)}\left[ 1 - 2(2 d_m - 1)\ln\left(\frac{z}{4}\right) \right] \, ,
\end{equation}
where $z(\alpha)$ is as defined in eq.\rref{defz}; from the
definition \rref{correlength} of the correlation length $l_c$, the
characteristic decay distance for the entanglement is then
$l_c/4$.    As the coupling strength is then increased, the
degeneracy is lifted, with the innermost modes showing the
greatest relative splitting.

This ``wave" of mode distortion and degeneracy lifting continues
from the inside of the chain out as the coupling increases until
the correlation length becomes  comparable to the size of the
block and the outermost modes are distorted.  As the coupling
strength is increased beyond this point, no appreciable change in
the shape of the modes is detected.  Thus  a critical value of
$\alpha$ determined by
\begin{equation}
 l_c(\alpha) \simeq N_b \,
\end{equation}
sets a threshold beyond which the mode shape structure becomes
frozen in its strong coupling configuration. It is this condition
that underlies the transition between the type $I$ and the type
$II$ transition in Fig. \ref{fig:entphases}.

\begin{figure}
\epsfxsize=2.8 truein  \centerline{\epsffile{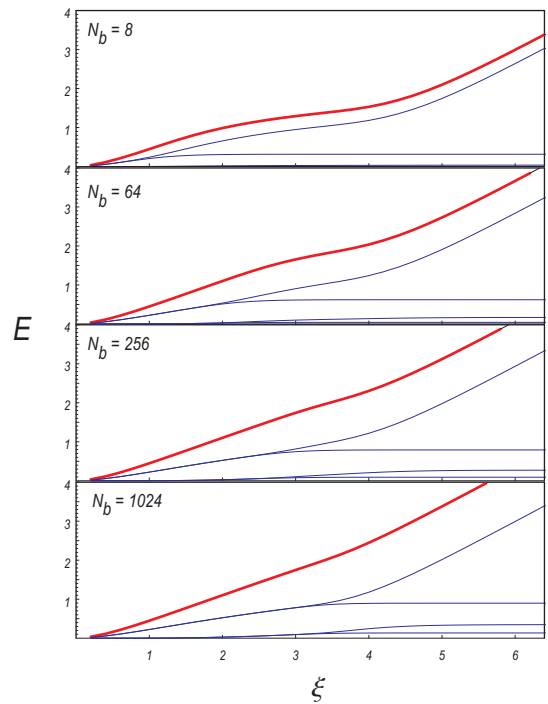}}
\medskip
\caption[\, ]{Contribution of the first four dominant modes (thin
lines) to the total entanglement (thick  line) as a function of
coupling strength $N_b$ for different values of the the block size
($N = 2048$). }\label{fig:modesalphaent}
\end{figure}

\begin{figure}
\epsfxsize=3.4truein  \centerline{\epsffile{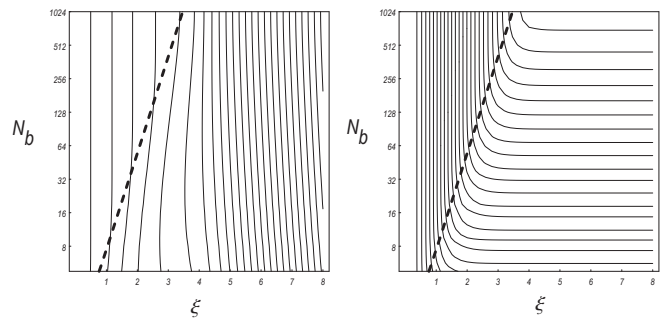}}
\medskip
\caption[\, ]{Level curves of entanglement contribution from first
(left) and second (right) Williamson modes to the total
entanglement depicted in Figures \ref{fig:3dtotent} and
\ref{fig:entphases}. The dotted line shows the correlation length
$l_c$, Eq. \rref{correlength}, as a function of coupling
strength.}\label{fig:2modescontr}
\end{figure}

The onset of this transition and the ensuing behavior of
entanglement thereafter is best appreciated from Fig.
\ref{fig:modesalphaent}, where we plot, as a function of the
coupling, the total entanglement  and the partial entanglement
from the first four dominant modes. The main signature of the
transition  is the lifting of the degeneracy involving the first
(even) and second (odd) modes, which together up to that point are
the predominant contributors to the total entanglement. Beyond
that point, however,  a clear decoupling occurs between the first
mode and the remaining modes: the first mode accounts for the
behavior of the entanglement as a function of coupling,
reproducing the three-regime curve (Fig. \ref{fig:singlent})
observed in section \rref{singlemode} for the case of the single
entangled oscillator; on the other hand, the remaining modes
become frozen in their coupling strength behavior. That this
``freezing out" indeed occurs when  $l_c(\alpha) \simeq N_b$ is
best seen from Fig. \ref{fig:2modescontr}, which shows the level
curves on the $\xi-N_b$ plane for the entanglement of the first
and second modes, together with a graph of the line $l_c(\alpha) =
N_b$.

In this way we find that  for strong coupling, the behavior of
entanglement as a function of $\alpha$, particularly with respect
to the type $II$ to type $III$ transition, is entirely due to the
single outermost ($m=1$) Williamson mode of the block. As observed
from the mode shape profiles (Fig. \ref{fig:uvmodes}), this mode
involves an essentially constant participation away from its
turning points (located at the edge of the block) and may
therefore be interpreted as a remnant of the $\theta_k=0$
collective normal eigenmode of the whole chain. In the context of
the strong regimes, we shall therefore refer to this mode as the
\textit{collective mode}; similarly, we shall use the term
\textit{residual modes} for the remaining Williamson modes of the
block. The reader is referred to sections \rref{stronganalcollect}
and \rref{stronganalresid} respectively for  further analysis of
the collective and residual behaviors.

\begin{figure}
\epsfxsize=2.8 truein  \centerline{\epsffile{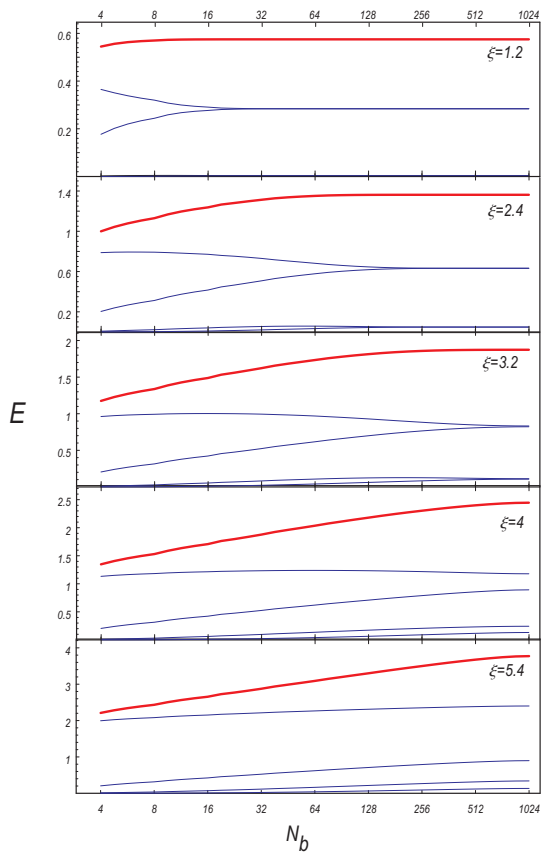}}
\medskip
\caption[\, ]{Contribution of the first four dominant modes (thin
lines) to the total entanglement (thick  line) as a function of
$N_b$ for different values of the coupling strength ($N = 2048$).
}\label{fig:modesnbent}
\end{figure}

Addressing finally  the large block size and chain size behavior
of entanglement, we find that in contrast  to the  large coupling
behavior, the responsible modes for the  $\sim \ln N_b$ behavior
are the residual modes. This can be seen from Fig.
\ref{fig:modesnbent}, which shows both the total entanglement and
the contribution from the first four dominant modes as a function
of $\ln N_b$ for various coupling strengths. For large enough
coupling, we find that the collective mode  shows a ``freezing" in
its $N_b$ behavior, while the greatest variation with $N_b$ is
shown by the first couple residual modes. A more careful
examination of the large $N_b$ behavior, however, indicates that
the entanglement contribution of  the outermost modes grows slower
than $\ln N_b$--rather like $\ln \ln N_b$. Thus, the $\ln N_b$
behavior of the total entanglement has to be attributed to a
cumulative effect from a certain number of residual modes that
contribute significantly in the large $N_b$ limit, which turns out
to be of order $\ln N_b$. If the modes are labelled by their
characteristic frequencies, this phenomena translates to an
enhancement by a factor  $\ln N_b$ of the density of states at
zero frequency\cite{peschel04}, and corresponds in the black-hole
analogy to the divergence of the density of states outside the
horizon in the absence of a UV cutoff\cite{callan}.

\begin{figure}
   \epsfxsize=3.4truein  \centerline{\epsffile{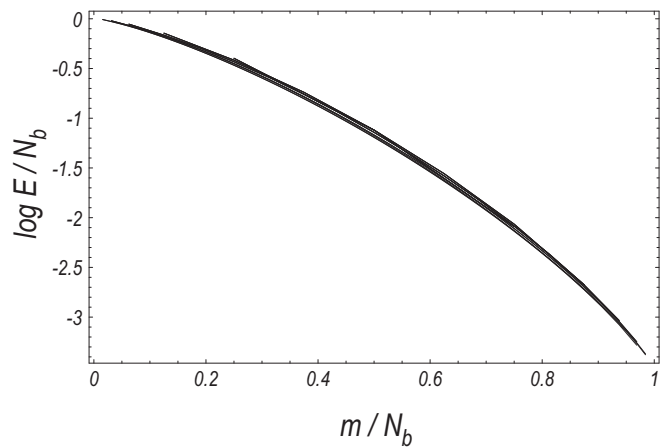}}
\medskip
\caption[\, ]{Scaling curve of entanglement versus mode number for
the strong regimes, for $N_b \ll N/2$. The plot includes data
taken from a chain of size $N=256$ ($N_b$ ranging from $4$ to
$32$), and chains of size $N=1024$ and $N=4096$ ($N_b$ ranging
from $4$ to $64$), all using $\xi = 10$ ($N_c \sim 10^7$).
}\label{fig:scaling}
\end{figure}

In this respect, it is instructive to look at  the entanglement
contribution of all the residual modes, for which $N_b \ll N/2$ in
the limit $\alpha \rightarrow 1$. As shown in Fig
\rref{fig:scaling}, the logarithm of the entanglement of the
$m$-th mode shows in this limit a scaling behavior
\begin{equation}\label{scalrel}
\ln E_m(N_b) \sim - N_b f\left(\frac{m}{N_b}\right)
\end{equation}
where $f(x)$ is some nonlinear function proportional to the
central frequency of oscillation of the mode.   Now, the
significant contribution to the entropy comes from modes for which
$\ln E_m$ is of magnitude unity or smaller, that is,  $ f \lesssim
\frac{1}{N_b}$.  As we then show in section
\rref{stronganalresid},  the function $f$ behaves for small values
of its argument like
\begin{equation}
-f(x) \ln f(x) \sim x \, .
\end{equation}
This behavior implies that an outer layer of residual modes, with
mode numbers  $m  \lesssim \ln N_b$, yield the relevant
contributions to the entropy.

To summarize the results of our qualitative survey of the
bipartite entanglement of  the harmonic chain, we emphasize again
the three-regime framework depicted in Fig. \ref{fig:entphases}.
We  observe a weak coupling regime (type $I$), characterized by
short range correlations,  well localized modes of definite
parity, and degenerate entanglement contribution from even and odd
combinations of oscillator sites at the same distance from the
edge. When the correlation length becomes of the order of the
chain size, the mode shapes acquire a more collective behavior, a
characteristic wavelength is established  for each mode, and the
degeneracy between contiguous modes of opposite parity is lifted.
A distinction also emerges between the outermost, or
``collective", mode and the remaining (``residual") modes. The
collective mode shows a minimal contribution to the large chain
size entanglement; however, it accounts for all the strong
coupling constant behavior thereafter, including the transition
between the type $II$ and type $III$ regimes. Conversely, the
residual modes become frozen  in their large coupling entanglement
contribution, but are responsible for the large chain behavior.
This behavior is conjectured to come from  a  cumulative effect
from a layer of the first  $\sim \ln N_b$ residual modes.

\section{Mode structure: analytic results}

We complement in this section the qualitative survey of the
previous section with a more careful  treatment of some of the
results presented there.  Our main aim is to produce simple
analytic models  that capture the essential elements of the
modewise entanglement structure in  weak and strong coupling.

\subsection{Entanglement in Weak coupling}
\label{weakanal}

As shown in our qualitative survey, entanglement for weak coupling
involves modes that are definite parity superpositions of local
oscillator-site modes. It will therefore be convenient to briefly
discuss the role of parity in our problem.

Parity selection rules follow from the fact that while the  local
correlation matrices $(G_A, H_A)$ and $(G_B, H_B)$  lose the
translational symmetry of the corresponding matrices for the whole
chain, they still possess a symmetry with respect to reflections
about the central indices, i.e., $G{}_{N_b + 1-l,N_b + 1-l} =
G{}_{N_b + 1-l,N_b + 1-l}$. Thus, the $G$ and $H$ matrices can be
written in block diagonal form as $G_A{} = G_A{}^+\oplus G_A{}^-$
and $H_A{} = H_A{}^+\oplus H_A{}^-$, in terms of blocks acting on
subspaces of  definite parity.  Similarly, it is easy to show that
with respect to the mode-mapping between the Williamson modes of
the block and those of its its complement, it is also possible to
show that the matrices $G_{AB}$ and $H_{AB}$ map modes of a given
parity definite parity in the block to modes of the same parity in
the complement.  Thus, the whole problem of finding the Williamson
modes in the size $N_b$ block, and their correlated counterparts
in the complement can be reduced to two separate problems of
finding Williamson modes for the sectors of size $N_b /2$
involving covariance matrices of definite parity.

>From Fig. \ref{fig:weakspecs} it is evident that for small values
of $\alpha$ the entanglement contribution decays exponentially
with the mode number. This leading order exponential behavior may
be understood  from simple arguments based on the ``wedge" shape
in Fig. \ref{fig:allmodes}. If for a given entangled mode-pair the
(properly-normalized) mode functions  $u_A^{(m)}$, $v_A^{(m)}$,
$u_B^{(m)}$ and $v_B^{(m)}$ are known, then  the respective
symplectic eigenvalue satisfies
\begin{equation}
\lambda_m^2 = v_A^{(m)}{}^T H_{A}G_A u_A^{(m)} =
\frac{1}{4}-v_A^{(m)}{}^T H_{AB}G_{BA} u_A^{(m)}\, .
\end{equation}
Using the resolution of the identity $\openone_B = \sum_\beta
v_B^{(\beta)}u_B^{(\beta)}{}^T$  this yields
\begin{equation}
\lambda_m^2 =   \frac{1}{4}-\sum_\beta v_A^{(m)}{}^T
H_{AB}v_B^{(\beta)}u_B^{(\beta)}{}^T G_{BA} u_A^{(m)}\, .
\end{equation}
However, we recall that $G_{BA}$ maps $u_A^{(m)}$  to $v_B^{(m)}$,
meaning that the only term surviving in the sum is the one in
which $\beta = m$. Thus we have that
\begin{equation}\label{lamformmodefs}
\lambda_m^2 =   \frac{1}{4}- v_A^{(m)}{}^T
H_{AB}v_B^{(m)}u_B^{(m)}{}^T G_{BA} u_A^{(m)}\, .
\end{equation}
Now, in the weak coupling regime,  we approximate the $m$-th even
or odd mode  functions in the weak regime by symmetric or
antisymmetric combinations of localized site positions of depth
$d_m$, where $d_m = m/2$ for $m$ even and $(m+1)/2$ for $m$ odd,
as suggested by Fig. \ref{fig:allmodes}:
\begin{eqnarray}
u_A^{(m\pm)}{}_i = v_A^{(m\pm)}{}_i & = &
\frac{1}{\sqrt{2}}[\delta_{i,d_m} \pm \delta_{i,N_b -d_m +1}]\\
u_B^{(m\pm)}{}_i = v_B^{(m\pm)}{}_i & = &
\frac{1}{\sqrt{2}}[\delta_{i,N_b+d_m} \pm \delta_{i,N-d_m+1}] \, .
\end{eqnarray}
Then, using the symmetries of the correlation functions Eq.
\rref{lamformmodefs} yields
\begin{equation}\label{lambadwedge}
\lambda_\alpha^2 = \frac{1}{4} - \left[\corP{N_b}^{(N)} \pm
\corP{2d_m-1}^{(N)}\right]\left[\corQ{N_b}^{(N)} \pm
\corQ{2d_m-1}^{(N)}\right]\, .
\end{equation}
Next, we make use of the  weak limit expressions for the
correlation functions, namely  that  both $g_l$ and $h_l$   behave
as $\sim z^l$, with $h_l$ negative for $l \geq 1$. Since $2d_m -1
< N_b$ always, then the leading order expansion of $\lambda
-\frac{1}{2}$ is expected to behave as $z^{2(2d_m-1)}$. Thus, in
this  approximation,  the symplectic eigenvalue is seen as being
due to  correlations between a site in the block at a depth $d_m$
from the interface and a ``mirror image" site in the complement at
the same distance to the interface.

A careful symbolic computation of this  leading order behavior for
small $N_b$ and assuming $N =\infty$ shows that in fact the
leading-order behavior of the symplectic spectrum is given by
\begin{equation}\label{appsimpspec}
\lambda_m(z) - \frac{1}{2} =
 \left(\frac{z}{4}\right)^{2(2 d_m - 1)} + o(z^{4d_m-1}) \, .
\end{equation}
This leads  to a mainly  exponential dependence with logarithmic
corrections expressed in eq. \rref{weakent} for the entanglement
per mode in the weak coupling regime, and yield the  solid lines
in Fig. \ref{fig:weakspecs}, showing
 excellent agreement for weak coupling up to
values of $z \simeq 0.7$. Our simple argument also leads us to
expect from Eq. \rref{lambadwedge} that the degeneracy between
even and odd states in Eq. \rref{appsimpspec} should be lifted by
an eigenvalue splitting of order $z^{N_b + 2m -1}$,  thus becoming
more significant as the mode-depth increases, and with the even
parity modes having the slightly larger symplectic eigenvalue.
This qualitative behavior is in fact  verified in Fig.
\ref{fig:weakspecs}.

\subsection{Strong coupling--The collective mode}
\label{collectivemode} \label{stronganalcollect}

We turn next to the emergence of the collective mode in the strong
coupling limit  and its entanglement behavior. This has to do with
the fact that the scale of the correlation function $g_l(\alpha)$
diverges as $\alpha \to 1$, together with the fact that the
function behaves logarithmically as a function of $l$. Thus, the
correlation function $g_l$ can  be separated as
\begin{equation}
g_l = g_0(\alpha) + \Delta_l\, ,
\end{equation}
where  $g_0$ is the diverging in $\alpha$ self-correlation
function, and $\Delta_l$ is tends to fixed limit independent of
$\alpha$. In turn, the correaltion matrix $G_A$ of the block takes
the form
\begin{equation}
   G_A \rightarrow N_b g_0\,  \chi \chi^T  + \Delta G \, ,
\end{equation}
where $\Delta G \sim O(1)$ and independent of $\alpha$, and
\begin{equation}\label{defu}
  \chi \equiv \frac{1}{\sqrt{N_b}}\left(%
\begin{array}{ccccc}
  1 & 1 & \ldots & 1 & 1 \\
\end{array}%
\right)^T \, .
\end{equation}
Note that since for fixed $N_b$,
\begin{equation}
\lim_{g_o \rightarrow \infty}  G_A \chi  \rightarrow N_b g_o [\chi
+  o(g_o^{-1}) ] \, ,
\end{equation}
the vector $\chi$ becomes an eigenvector of $G_A$ in that limit
with eigenvalue $N_b g_0$.

Similarly,  since the  momentum correlation functions are regular
and tend to a fixed value as $\alpha \rightarrow 1$,
 the product $H_A G_A$ is also split into diverging and finite parts:
\begin{equation}
H_A G_A = N_b\, g_0 (H_A \chi)(\chi^T) + H_A \Delta G\, .
\end{equation}
We now construct the vectors
\begin{equation}
u_c \propto H_A \chi , \, \ \ \ \ \ v_c \propto  \chi , \, ,
\end{equation}
with normalization set so that $u_c \cdot v_c = 1$, and define
\begin{equation}
g_\chi \equiv \chi^T H_A \chi = N_b g_0 \, ,\ \ \ \ \ h_\chi
\equiv \chi^T H_A \chi \, .
\end{equation}
Then, we have that
\begin{eqnarray}
\lim_{g_o \rightarrow \infty} H_A G_A u_c & \rightarrow  & g_\chi h_\chi [ u_c + o(g_o^{-1})] \\
\lim_{g_o \rightarrow \infty} v_c^T H_A G_A  & \rightarrow  &
g_\chi h_\chi [ v_c^T + o(g_o^{-1})] \, ,
\end{eqnarray}
showing that $u_c$ and $v_c$ become the right and left
eigenvectors of $G_A H_A$ in the strong coupling limit with
symplectic eigenvalue
\begin{equation}
\lambda^2_c =  g_\chi h_\chi \, .
\end{equation}
Next, we compute the value of $h_\chi$,
\begin{eqnarray}
h_\chi & = & \chi^T H \chi \nonumber \\
& = & \frac{1}{N_b}\sum_{i=1}^{N_b}\sum_{j=1}^{N_b} \left[
-\frac{\sqrt{2}}{\pi}\frac{1}{4(i-j)^2 -1}\right] \nonumber \\
& = & \frac{1}{\sqrt{2} N_b \pi}\left[\psi\left(N_b +
\frac{1}{2}\right) + \ln(4) + \gamma \right] \, ,
\end{eqnarray}
where $\psi$ is the DiGamma function.  For large $N_b$, we
approximate
\begin{equation}
h_\chi \simeq \frac{1}{\sqrt{2} N_b \pi}\ln\left(4N_b\right) \, ,
\end{equation}
so that
\begin{equation}
\lambda_c^2 \simeq \frac{g_o}{\sqrt{2}  \pi}\ln\left(4N_b\right)
\,
\end{equation}
becomes the symplectic eigenvalue the approximation.

The behavior of the collective mode entanglement for the type-$II$
and type-$III$ regimes follows the same analysis performed in
section \ref{singlemode} for the single entangled oscillator,
based on the two regimes of strong coupling behavior of $g_o$
determined by $N_t(\alpha)$ and $N_c(\alpha)$. Asymptotically, we
find that
\begin{equation}\label{singlentapps}
E \sim \left\{ \begin{array}{cc}
  \frac{1}{2}\ln \ln  N_b+ \frac{1}{2}\ln \ln  N_t(\alpha)\ \, ,  &  \   N_t(\alpha) \sim N \gtrsim N_c \\
{} & {} \\
\frac{1}{2}\ln \ln  N_b + \frac{1}{2}\ln N_t(\alpha)/N  \ \, , & \
N_c(\alpha) \gg N\ \ \  \, .
\end{array} \right.
\end{equation}
Thus, the entanglement curve of the collective mode is essentially
that of the single entangled oscillator, except for sublogarithmnc
corrections  dependent on  the chain size $N_b$.

\subsection{Strong Coupling--Residual Modes}
\label{stronganalresid}

General  qualitative and quantitative aspects of residual mode
entanglement in the strong regime can be illustrated through a
simple analytical model in the continuum, along the lines of
similar models discussed in the context of geometric entropy in
black hole physics \cite{callan} and reduced density matrices for
free electron chains \cite{peschel04}.  Such models are
 useful in deriving the correlation between mode
number, wavelength and turning point location, and thus can
account for the  scaling relation depicted in Fig
\ref{fig:scaling}, and the density of states determining the $\ln
N_b$ behavior.

As discussed in section \ref{gaussect},  an eigenvalue problem
equivalent to $H G u= \lambda^2 u$ for a given region $A$, is the
eigenvalue problem
\begin{equation}
 C u^{(m)} = - \kappa_m^2 u_A^{(m)} \, , \ \ \ C \equiv H_{AB}
 G_{AB}^T\, ,
\end{equation}
where $H_{AB}$ and  $G_{AB}$ are the  matrices containing
correlations between the two complementary regions $A$ and $B$,
and where the eigenvalue  $\kappa_m$ is related to the symplectic
eigenvalue $\lambda_m$ according to $\lambda_m^2 = \frac{1}{4} +
\kappa_m^2\, $. We assume henceforth an infinite chain, and for
the block adopt an index convention  centered at the block
midpoint, i.e., so that indices run from $-(N_b-1 -1)/2$ to
$(N_b-1 -1)/2$; no loss of generality is entailed by assuming
$N_b$ odd so that indices are integers. With these assumptions,
the matrix elements of $\Gamma$ are given by the sums  to infinity
over both regions of the complement bordering the block
\begin{equation}\label{gammasum}
C_{i j} =\sum_{k=\frac{N_b +1}{2}}^{\infty} h_{k-i} g_{k-j} +
\sum_{k=\frac{N_b +1}{2}}^{\infty} h_{k+i} g_{k+j} \, .
\end{equation}
We base our computation of $C$ on the asymptotic  expressions for
the correlation functions \rref{gasymp} and \rref{gasymp},
retaining the terms yielding the leading order behavior in $N_b$
upon converting the above sums to integrals. Thus we use
\begin{eqnarray}
g_{i -j} & \simeq & g_0 - \frac{1}{ \sqrt{2 }\, \pi} \ln |i -
j|\, , \nonumber \\
h_{i -j} & \simeq & - \frac{1}{ 2 \sqrt{2}\, \pi} \frac{1}{|i -
j|^2} \, .
\end{eqnarray}
where we include the diverging part $g_0$ corresponding to the
collective mode. Replacing the sums in \rref{gammasum} by
integrals in $k$ from $N_b/2$ to infinity,  and defining  the
scaled variables $x \equiv i/N_b$, and $y \equiv i/N_b$, we
finally obtain a result of the form
\begin{equation}
C(i,j) = N_b^{-1}\Gamma(x,y)\, , \ \ \ \ \Gamma= \Gamma_{CS} +
\Gamma_{CA} + \Gamma_{R}\, , \ \ \
\end{equation}
where
\begin{eqnarray}\label{gammas1}
\Gamma_{CS} & = & -\frac{\sqrt{2}}{ 4 \pi  (\frac{1}{4} - y^2)}
\left[ g_0 - \frac{\ln(N_b \sqrt{\frac{1}{4} - x^2})}{\sqrt{2} \pi} \right] \nonumber \\
\Gamma_{CA} & = & \frac{x}{ 4 \pi^2  (\frac{1}{4} - x^2)}\, \ln \left( \frac{1 + 2y}{1- 2 y} \right)\nonumber \\
\Gamma_{R} & = & \frac{1}{ 4 \pi^2  (x - y)} \ln \left( \frac{(1 +
2x)(1- 2 y)}{(1- 2 x)(1+ 2 y)} \right)\, .
\end{eqnarray}
The matrices $\Gamma_{CS}$ and $\Gamma_{CA}$ are of rank one and
determine the behavior of the dominant mode  for the even and odd
sectors respectively; in turn, the bulk of the residual modes is
determined by $\Gamma_{R}$.

To probe the solutions to the above eigenvalue problem, we next
consider the continuum generalization,
\begin{equation}\label{evprocont}
\int_{-1/2}^{1/2} dy \, \Gamma(x,y)\, \psi(y) = - \kappa^2
\psi(x)\, .
\end{equation}
For this, it will be convenient to perform a change of variables
suggested by \rref{gammas1}, mapping the interval $(-1/2, 1/2)$ to
the real line
\begin{equation}
u = \ln \left( \frac{1 + 2x}{1- 2 x} \right) \, , \ \ \ v = \ln
\left( \frac{1 + 2y}{1- 2 y} \right)\, ,
\end{equation}
and  rescale the wave function  according to
\begin{equation}
\tilde{\psi}(u)  =  \frac{1}{4}\cosh^{-2}\left(
\frac{u}{2}\right)\, \psi(u)\, .
\end{equation}
The new eigenvalue equation then reads
\begin{equation}
\int_{-\infty}^{\infty} dv \, \tilde{\Gamma}(u,v)\, \psi(v) = -
\kappa^2 \psi(u)\, ,
\end{equation}
where the new kernel $\tilde{\Gamma}(u,v)$ is split in the form
$\tilde{\Gamma}_{CS} + \tilde{\Gamma}_{CA} + \tilde{\Gamma}_{R}$,
with the respective terms  given by
\begin{eqnarray}\label{gammas1}
\tilde{\Gamma}_{CS} & = & -\frac{\sqrt{2}}{ 4 \pi}g_0 - \frac{1}{4
\pi^2}\left[  \ln\left(\frac{N_b}{2 \cosh\left(
\frac{v}{2}\right)} \right) -\frac{u}{2} \tanh\left(
\frac{u}{2}\right) \right] \nonumber \\
\tilde{\Gamma}_{CA} & = & \frac{v}{ 4 \pi^2 }\tanh\left(
\frac{u}{2}\right)\nonumber \\
\tilde{\Gamma}_{R} & = & \frac{1}{ 8 \pi^2 } \, \frac{u - v}{
\tanh \left( \frac{u-v}{2}\right)}\, .
\end{eqnarray}
We now use $\tilde{\Gamma} \simeq  \tilde{\Gamma}_R$ for the inner
modes, which amounts to neglecting  the lowest wavelength
solutions. In this case, it is straightforward to see that the
plane wave function $\tilde{\psi}(u) = \exp(i \omega u )$ is an
eigenfunction of $\tilde{\Gamma}_R$. The resulting integral for
the eigenvalues can be performed by contour methods and yields
\begin{equation}
\kappa_R^2 = -\frac{1}{8 \pi^2} \int_{-\infty}^{\infty} dv
\frac{v\, e^{ i \omega v}}{ \tanh \left( \frac{v}{2}\right)}\,   =
\frac{1}{4}\sinh^{-2}(\pi \omega)\, .
\end{equation}
Finally, reverting  to the variables $x$, $y$  and taking even/
odd combinations of the plane wave solutions, we find as finite-
wavelength eigenfunctions of the continuum eigenvalue problem
\rref{evprocont}, the solutions
\begin{equation}
\psi(x) = \frac{1}{1  - 4 x^ 2}\left\{ \begin{array}{c} \cos(
S_\omega(x) ) \\ \sin( S_\omega(x) ) \end{array} \right\}
\end{equation}
where
\begin{equation}\label{defs}
S_\omega(x) = \omega \ln \left( \frac{1 + 2x}{1- 2 x} \right)\, ,
\end{equation}
The local wavenumber of these solutions, $S'_\omega(x)$, increases
from the value $4 \omega$ as one moves away from the midpoint  and
diverges at the boundaries ($x = \pm 1/2$). Similar mode functions
were obtained in \cite{peschel04} for the eigenmodes of the
reduced density matrix of the free-electron chain.   Note that the
symplectic eigenvalues are
\begin{equation}
\lambda = \sqrt{\frac{1}{4} + \frac{1}{4}\sinh^{-2}(\pi \omega) }
= \frac{1}{2}\coth(\pi \omega) \, ,
\end{equation}
so the  Boltzmann-like $\beta_m$ factors associated with each
Williamson mode according to eqs. \rref{sim-beta} are given by $
\beta = 2 \pi \omega$.  For large $\beta$, this yields an
asymptotic expansion of the entropy $E \sim e^{-2 \pi \omega}$,
from which the scaling relation depicted in Fig. \ref{fig:scaling}
should then be expected to follow.

\begin{figure}
   \epsfxsize=3.5truein  \centerline{\epsffile{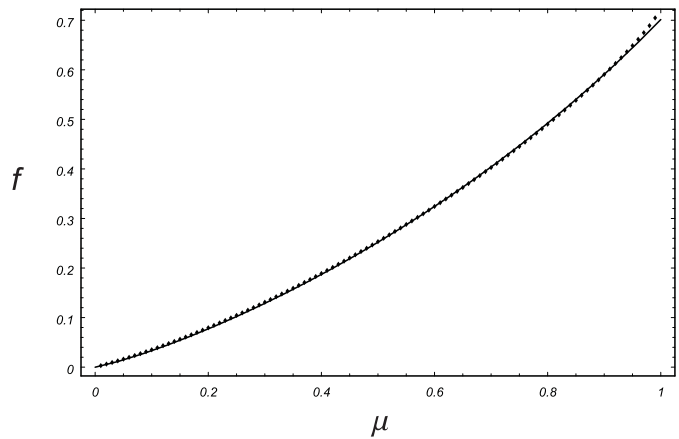}}
\medskip
\caption[\, ]{Comparison between  analytical and numerical results
for the scaled central wavelength $f$ as a function of  the scaled
mode number, using $\zeta = 0.45$. }\label{fig:scalecomp}
\end{figure}

We next connect these solutions to the corresponding solutions of
the discrete chain. The first thing to note is that in the
continuum approximation, the discrete modes correspond to averages
of the continuum modes over the lattice spacing. Thus, the
diverging oscillatory behavior of the continuum modes as  the
block edge is approached holds only up to a certain distance from
the edge in the discrete approximation, corresponding to the point
where the local oscillation wavenumber becomes of the order of the
cutoff imposed by the lattice spacing. As one moves towards the
edges beyond this point, the oscillations are washed out in the
average and  thus the mode amplitude decays. This is consistent
with the fact, noted in section \ref{qualitative}, that at the
turning points each Williamson mode shows oscillations of the
maximum wavenumber. The turning points are therefore fixed by the
condition that
\begin{equation}
\zeta\left|\frac{dS}{dx}\right|_{x = \pm x_t} =  \pi N_b
\end{equation}
where $\zeta$ is number of order unity. With the definition
$\rref{defs}$, this yields  the relation between the turning point
and $\omega$
\begin{equation}
x_t = \frac{1}{2} \sqrt{1 -  \zeta \frac{4 \omega}{\pi N_b}  } \,
.
\end{equation}
Once the turning point is identified as a function of $\omega$, we
can work out the quantization of the modes. For this we use the
fact observed from our numerical calculations in section
\ref{qualitative}, that when the interior modes are modulated by
the oscillating factor $(-1)^i$--corresponding to frequency shift
by the cutoff frequency--the modes show a hierarchy in which each
mode can be labelled by the number of nodes $n+N_b - m$. Writing
the modulated functions like
\begin{equation}
\psi_{\mathrm{mod}}(x) = (-1)^i \psi(x) =\frac{1}{1  - 4 x^
2}\cos(\pi N_b x - S_\omega(x) )\, ,
\end{equation}
and  similarly for the antisymmetric mode, we demand that
$\psi_{\mathrm{mod}}(x)$ show $n$ nodes between the turning
points; in this way we obtain the quantization condition
\begin{equation}\label{quantcond}
\pi N_b x_t(\omega) - S_\omega(x_t(\omega)) = \frac{n \pi}{2} \, .
\end{equation}
The resulting equation for $\omega$  is best cast in a form that
can easily be interpreted from the scaling relation of the entropy
\rref{scalrel}. For this we define the dimensionless parameters
\begin{equation}
f \equiv \frac{4 \omega}{\pi N_b} \, \ \ \ \  \mu = \frac{m}{N_b}
\, ,
\end{equation}
where  $m = N_b - n$ labels the modes from the outside in, and $f$
measures the central wavelength of the mode relative to the cutoff
wavelength $\pi N_b$; for large $\omega$, the entanglement of the
mode is therefore given by
\begin{equation}
N_b^{-1} \ln E = \frac{\pi^2}{2} f(\mu) \, ,
\end{equation}
in accordance with equation \rref{scalrel}. From the quantization
condition \rref{quantcond}, the function $f(\mu)$ is then the
solution to the equation
\begin{equation}
1 - \sqrt{1 - \zeta f} + \frac{f}{2}\ln\left(\frac{1 + \sqrt{1 -
\zeta f}}{1 - \sqrt{1 - \zeta f}}\right) = \mu \, .
\end{equation}
For the  outermost modes, corresponding to small $\mu$, small $f$,
the relation becomes independent of $\zeta$ to leading order and
reduces to
\begin{equation}\label{smallf}
f \ln f = - 2\mu \, ,
\end{equation}
as mentioned earlier. On the other hand, the fit for the innermost
modes is sensitive to the precise value of $\zeta$. We have found
that a value of $\zeta \simeq 0.45$ yields a remarkably good fit
to the numerical data (Fig \ref{fig:scalecomp}), as well as the
turning point location depicted in Fig. \ref{fig:turningpoint}.

It is worth mentioning that in the present approach, the role of a
cutoff frequency for the continuum is not only reflected  in a
cutoff of the resulting frequency spectrum of the reduced density
matrix, but also entails a \emph{localization} of the mode
functions away from the interface between the block and the
complement. The condition determining the turning point can only
be approximately estimated from the continuum model and therefore
fails to account for the exponential fall-off of the amplitude
beyond the turning point. The method also fails to account for the
$ \Delta x \sim N_b^{-1/2}$  scaled width of the the innermost
mode functions , which is sensitive to the  fall-off details.

\subsection{Residual mode contribution to the Entropy}
\label{asymptotic}

To obtain the asymptotic $\ln N_b$ behavior for $N_b \ll N$, we
note that the significant contributions to the entropy will come
from modes for which $\omega$ is of order unity or smaller, since
the entanglement  is suppressed exponentially for $\omega > 1$;
since this implies that $f$ will be of order $1/N_b$ or smaller,
the approximation \rref{smallf} becomes more accurate for these
modes as  $N_b \rightarrow \infty$. Thus,  discarding terms of
order unity in the logarithm, the eigenvalue equation for $\omega$
can be cast in the implicit form
\begin{equation}\label{omegaouter}
\omega_m  = \frac{m \pi}{2 \ln(N_b)\left[ 1 - \frac{\ln
\omega}{\ln N_b}\right]} \, .
\end{equation}
Next, we can argue that as $N_b \rightarrow \infty$, the
 term in brackets in the denominator of \rref{omegaouter}can be
neglected for the calculation of the entanglement; this term
becomes important when  $N_b^{-1} \sim \omega \ll 1$ in the lower
(since  $\omega \gtrsim 1$, the approximation is already warranted
in the upper end). For the lower end, we consider the smallest
posible value of $\omega$, for $m = 2$; according to
\rref{omegaouter}, this value is of order of $1/\ln N_b$ and is
therefore of an order $N_b/\ln N_b$ greater than the scale at
which \rref{omegaouter} deviates from linearity. Hence we find the
approximate linear expression for $\omega$,
\begin{equation}
\omega_m \simeq \frac{m \pi}{2 \ln(N_b)} \, , \label{omegalin}
\end{equation}
for the  range of modes yielding relevant contributions to the
entanglement as $N_b \rightarrow \infty$. Using the approximation
$E(\beta) \simeq 1 - \ln \beta$ for $\beta \ll 1$, the
entanglement of the modes  for which $ \ln N_b \gg m \geq 2 $, can
then be estimated to be
\begin{equation}
E_m \simeq - \ln \left( \frac{\pi^2 m }{ \ln N_b} \right) \, .
\label{entlin}
\end{equation}
Consequently, the outermost residual modes yield a contribution of
order  $ \ln \ln N_b $, in the same way that the collective mode
behaves for large $N_b$.

To obtain the leading term in the asymptotic expansion of the the
total residual mode entanglement, we note that   for $N_b \gg m
\gg \ln N_b$ the entanglement contribution is suppressed, the
total residual entanglement can be estimated to be
\begin{equation}
E_R \simeq \left. \left(\frac{d m}{d \beta}\right)\right|_0
\int_{0}^{\infty} d\beta\, S( \beta)
\end{equation}
where
\begin{equation}
S(\beta) = \frac{\beta}{ e^{\beta} -1} - \ln(1 - e^{-\beta}) \, ,
\end{equation}
and where we evaluate the density of states assuming a linear
relation at $\beta = \omega =0$ (note that this approximation
entails  neglecting terms of order $o(\ln \ln N_b)$ and lower).
With an integration by parts, the entropy can expressed as
\begin{equation}
E_R = 2 \left(\frac{d m}{d \beta}\right) \int_{0}^{\infty}\,
d\beta \, \frac{\beta}{ e^{\beta} -1}  + O(\ln \ln N_b)\, ,
\end{equation}
and from \rref{omegalin}, the density of states is given by
\begin{equation}
\frac{d m}{d \beta} = \frac{1}{2 \pi}\frac{d m}{d \omega}=
\frac{1}{\pi^2}\ln N_b \, .
\end{equation}
The integral is standard in statistical mechanics and is given by
$\pi^2/6$. Thus we find that as $N_b \rightarrow \infty$, the
residual mode entropy is given by
\begin{equation}\label{logent}
E_R = \frac{1}{3}\ln N_b \, + O(\ln \ln N_b)\, ,
\end{equation}
as expected.

\section{Comparison with previous results}

In this section we comment on the relation of our results with
earlier work. Audenaert et. al. \cite{plenio} have studied the
entanglement in the circular linear chain model, for various
choices of bipartite divisions. In this work, the logarithmic
negativity \cite{log-negativity} has been used as a measure for
entanglement. Interestingly, we find that the von-Neumann
entanglement seems to be, in the present problem, a more sensitive
quantifier of the connection between entanglement and
correlations. In Section IIIa we showed that the behavior of
vacuum correlation functions quantifies three regimes: a weak
coupling regime characterized by short-ranged correlations, an
intermediate regime that is reached when the correlation length is
of the order of the whole chain, and finally, a long range
correlation regime. The transition between these regimes is
clearly manifested in the behavior of  the von-Neumann entropy
(see  Figs. 5, 7, 15), but is absent in the behavior of the
logarithmic negativity. The difference also shows up when
comparing the dependence of the two measures on the total chain
size $N$. In the particular case where $N_b=N/2$ (\cite{plenio},
corollary 1), the logarithmic negativity shows no $N$ dependence.
The von-Neumann entropy, on the other hand, decreases as a
function of $N$ like $\ln(N_t/N)$ (see eqs. (4.8), (6.21) and Fig.
9). This dependence on $N$, can be physically understood as due to
the contribution of the collective mode to the entanglement, which
reduces with increasing $N$. Thus is seems that the logarithmic
negativity is not sensitive to the contribution of the collective
mode.

The entanglement of a finite region for a one dimensional field
bosonic and fermionic fields has been previously investigated in
connection with the black hole entropy ``area law"
\cite{bombelli,srednicki,dowker}. By employing methods of
conformal field theory it has been shown
\cite{callan,holzhey,kabat} that in the massless case entanglement
behaves like \beq \label{central} {c+ \bar c \over 6} \ln {L\over
\epsilon} \eeq where $\epsilon$ plays the role of the UV cutoff,
and $c$ and $\bar c$ are the holomorphic and anti-holomorphic
central charges of the conformal field theory, with $c=\bar c =1$
for bosons, and $c=\bar c= 1/2$ for fermions. Thus the overall
coefficient is given by either 1/3 for bosons,
 or 1/6 for fermions.
The same type of universal behavior has been recently  derived
the XY and Heisenberg spin-chain models \cite{vidal}.

In the present work,  we obtained
a logarithmic dependance of the entanglement $\frac{1}{3}\ln N$,
 corresponding as expected to a bosonic field (Fig. \rref{fig:ln3}).
Furthermore we have seen that the logarithmic increase of the can
be understood as an increase in the number of relevant
contributing modes while the coefficient $1/3$ can be obtained
from the density of the modes. As in previous results, these modes
can be identified to be in a layer that becomes infinitesimally
narrow in the limit of large $N$. However, no attention has so far
been given to the structure of the inner modes. A central outcome
of the present work is that the inclusion of an ultraviolet
cutoff, which is needed for the consistency of the one dimensional
field theory, gives rise to a localization of the highest
frequency modes around the midpoint of the region. Although the
contribution of these modes to the entanglement is exponentially
small it is plausible that these inner modes play an important
role in the persistence of vacuum entanglement between separated
regions as we suggest in the next section.

A number of results presented in this paper can also be related
to previously obtained results for both fermions and bosons in the
context of the density matrix renormalization group
(DMRG)\cite{whiteDMRG,whiterev,peschel04,peschel98,peschel99,peschel00}.
In particular, the factorized thermal form of the reduced density
matrices, the shape of the corresponding mode functions in the
continuum limit, and the approximately linear behavior of the
frequency spectrum for the outermost modes have been studied
extensively in that context. It is possible that the method used
in Section \ref{stronganalresid}, whereby the cutoff imposed by
the lattice spacing is used to establish the turning points and
quantization condition of the mode functions, may prove useful for
the DMRG scheme.

\section{Summary and discussion}

Previous work on ground state entanglement in chain-like systems
has mostly focused on the dependence of the amount of entanglement
on parameters such as the block size, the separation between sites
and the nature of the bipartite splitting. While in the present
work we have reproduced several earlier results, as discussed in
the previous section, the emphasis here has been on the study of
the spatial entanglement structure emerging from the modewise
decomposition of the ground state wave function.

From this analysis, we have identified certain general properties
of the  mode structure and its relation to the entanglement
contribution. A first feature is \emph{localization}--a definite
characteristic distance from the division interface for the
entangled modes at either side of the interface, thus establishing
a characteristic distance separating   the entangled mode
functions. This  in turn, serves to characterize the strength of
the entanglement, which decays exponentially with this distance.

A second feature, which becomes sharper with increasing coupling
strength, is a characteristic wavelength correlated with the
degree of localization of the modes. This correlation is in fact
observed in two guises: on the one hand, it provides an
alternative characterization of the modes in terms of their
central wavelength, with the  innermost modes possessing the
shortest wavelength dictated by the lattice spacing; on the other
hand, the amplitude of each mode is correlated to the local
oscillation wavelength, with the largest amplitude occurring when
a the cutoff wavelength.

We have shown that the effect of the interaction strength on the
shape of the modes and their contribution to entanglement is
fundamentally connected with the correlation length. When the
coupling is strong enough such as the correlation length becomes
comparable to the size of the system, and the system becomes
effectively massless, scale free behavior emerges for the bulk of
the modes. The shape of the mode functions can be connected to the
scale free continuum field theory, and both the localization and
the characteristic wavelength scale with the size of the block.

On a more speculative note, it is possible the results of this
paper may shed new light into several features of mixed state
entanglement for separated non-complementary regions in vacuum. It
has been shown that for arbitrarily  separated regions, vacuum
entanglement persists and Bell inequalities are violated
\cite{vac-bell-ineq}, with a lower bound of the entanglement that
goes like $\exp(-L^2/D^2)$ where $D$ and $L$ denote the size of
the regions and their separation. It was shown that a large probe
energy gap is required in order to extract this entanglement. This
seems to suggest that the localization of the inner modes and
their  short wavelength characteristic are linked with the
persistence of vacuum entanglement at large distances. It is
possible this persistence represents an effective shielding of the
entanglement content of the innermost modes. This qualitative
argument could help explain the truncation effect that takes place
beyond a critical distance in the discrete version of this problem
when the region sizes are kept fixed, and could explain the
discrepancy between entanglement and correlation lengths in other
models.

\begin{acknowledgements}
A.B. acknowledges support from Colciencias (contract
1204-05-13622). B.R. acknowledges the support of ISF grant
62/01-1.
\end{acknowledgements}

\section{Appendix: Discretization of the Massive Continuum Theory}

In this  appendix we connect our results for the discrete chain with
a continuum bosonic theory in one dimension. To this end, consider
the Hamiltonian for the one-dimensional massive continuum theory
on a circular topology
\begin{equation}\label{hamcont}
    H = \frac{1}{2}\int dx \, \left[ \pi(x)^2 + (\phi'(x))^2
    + \mu^2 \phi(x)^2 \right] \, ,
\end{equation}
where we assume that $x$ runs from $-L/2$ to $L$ and that
$\phi(x-L/2) = \phi(x+L/2)$, and  field configuration and momentum
operators satisfying the commutation relations
\begin{equation}
[\phi(x,t), \pi(y,t) ] = i \delta(x-y) \, .
\end{equation}
The Hamiltonian is diagonalized in terms of normal mode creation
(annihilation) operators $a(k_n) (a(k_n)^\dag)$ with $k_n =
\frac{2 \pi }{L}n$, $n \in \mathbb{Z}$, and such that
$[a(k_n),a(k_m)^\dagger] = \delta_{n,m}$. The fields are then
given by
\begin{eqnarray}
\phi(x) & = & \frac{1}{\sqrt{ L}}\sum_n\,
\frac{1}{\sqrt{2\omega(k_n)}} \left[a(k_n) e^{i k_n x}\!
+\!\mathrm{h.c.}
\right] \, ,\\
\pi(x) & = & \frac{-i}{\sqrt{ L}}\sum_n
\sqrt{\frac{\omega(k_n)}{2}} \left[a(k_n) e^{i k_n x}\! -\!
 \mathrm{h.c.}\right] \, ,
\end{eqnarray}
where $\omega(k)$ is the usual dispersion relation $\omega(k) =
\sqrt{\mu^2 + k^2}$. The continuum field correlation functions can
then be obtained yielding
\begin{eqnarray}
\tilde{g}^{(L)}(x) & = & \frac{1}{2L}\sum_n \frac{\cos(k_n
x)}{\omega(k_n)} \, \\
\tilde{h}^{(L)}(x) & = & \frac{1}{2L}\sum_n \omega(k_n) \cos(k_n
x)
\end{eqnarray}
In the limit when $L \rightarrow \infty$, these expressions can be
expressed in terms of the modified Bessel functions:

\begin{eqnarray}\label{contginfty}
\tilde{g}^{(\infty)}(x) & = & \frac{1}{4
\pi}\int_{-\infty}^{\infty} dk \frac{\cos( k x)}{\sqrt{k^2 +
    \mu^2}}\\
    & = & \frac{1}{2 \pi}K_0( \mu |x| )\\
\tilde{h}^{(\infty)}(x) & = & \frac{1}{4
\pi}\int_{-\infty}^{\infty} dk \sqrt{k^2 +
\mu^2}\ \cos( k x)\\
 & = & - \frac{\mu }{2 \pi x}K_1( \mu |x| )
\end{eqnarray}

The asymptotic form of the correlation functions is given for  $x
\ll \mu^{-1}$ by
\begin{eqnarray}  \label{smallx}
\tilde{g}^{(\infty)}(x) & \rightarrow & -\frac{1}{2
\pi}\left[\ln\left(\frac{\mu|x|}{2}\right) + \gamma\right] \\
\tilde{h}^{(\infty)}(x) & \rightarrow & -\frac{1 }{2 \pi|x|^2}
\end{eqnarray}
and for $x \gg \mu^{-1}$, we have
\begin{eqnarray}
\tilde{g}^{(\infty)}(x)& \rightarrow & -\frac{e^{-\mu |x| }}{2
\sqrt{2\pi \mu |x| }} \\
\tilde{h}^{(\infty)}(x) & \rightarrow & \sqrt{\frac{\mu}{8 \pi
|x|^3} }e^{-\mu |x|} \label{largex}
\end{eqnarray}

The theory at the continuum may be approximated  by a linear chain
of $N$ sites in a ring topology,  introducing discretized
variables $q_n$ and $p_n$, which up to scale changes samples the
field and its conjugate momentum field at points $x_n = -L/2 +
\frac{n}{N} L$. To obtain a Hamiltonian of the form
\rref{hamchain1}, we first approximate the field Hamiltonian by
replacing $\int dx \rightarrow   \frac{L}{N}\sum_n $,  $
\phi'(x_n) \rightarrow \frac{N}{L}(\phi_n\!-\!\phi_{n-1})$. Then
we perform the following transformation
\begin{equation}
\phi(x_n) = \sqrt{\frac{N}{L}} \, \Lambda^{-1/2} \, q_n \, , \ \ \
\pi(x_n) = \sqrt{\frac{N}{L}} \, \Lambda^{1/2} \, p_n
\end{equation}
It is can now be seen that with the choice
\begin{equation}
\Lambda = \left[ 2 \left(\frac{N}{L}\right)^2 + \mu^2
\right]^{1/2}\, .
\end{equation}
 we obtain a Hamiltonian of the form \rref{hamchain1}
 with
 \beq E_o
= \Lambda \eeq
 and an effective coupling strength
\begin{equation}
\alpha(\mu L/ N) = \frac{1}{1 + \frac{1}{2}\left( \frac{\mu L}{N}
\right )^2 } = 1 - \left(\frac{\mu}{\Lambda}\right)^2\, .
\end{equation}

The correlation functions in the continuum are characterized by
the length scale $1/\mu$. As can be seen from eqs.
(\ref{smallx}-\ref{largex}),for separation $x < 1/\mu$ the theory
behaves essentially as massless, while for $x>1/\mu$, the
correlations decay exponentially. This length scale can then be
related to the correlation length obtained from the infinite
harmonic chain by the relation $\mu a \leftrightarrow N_c(\alpha)
= \sqrt{2\over {1-\alpha}}$. We can then verify that the following
relations hold between the discrete and continuum correlation
functions
\begin{eqnarray}
 \tilde{g}^{L}(x) & = & \frac{1}{\sqrt{ 2}}\lim_{N \rightarrow
\infty} \left.g^{(N)}_{n(x,N)} \right|_{\alpha(\mu L/ N) }\\
\tilde{h}^{L}(x) & = & \sqrt{2}\lim_{N \rightarrow \infty} \left(
\frac{N}{L}\right)^2\left.h^{(N)}_{n(x,N)} \right|_{\alpha(\mu L/
N)}
\end{eqnarray}
In finite massive chains, another useful limits are
\begin{eqnarray}
 \tilde{g}^{\infty}(x) & = & \frac{1}{\sqrt{ 2}}\lim_{\Lambda
\rightarrow \infty} \left.g^{(\infty)}_{n = x \Lambda/\sqrt{2}}
\right|_{\alpha(\mu/ \Lambda)} \\
\tilde{h}^{\infty}(x) & = & {\sqrt{ 2}}\lim_{\Lambda \rightarrow
\infty} \Lambda^2\left.h^{(\infty)}_{n = x \Lambda/\sqrt{2}}
\right|_{\alpha(\mu / \Lambda)}
\end{eqnarray}
or equivalently
\begin{eqnarray}
 \tilde{g}^{\infty}(x) & = & \frac{1}{\sqrt{ 2}}\lim_{l
\rightarrow \infty} \left.g^{(\infty)}_l
\right|_{\alpha=1-\frac{1}{2}\left(\frac{\mu x}{l}\right)^2} \\
\tilde{h}^{\infty}(x) & = & \sqrt{2}\lim_{l \rightarrow \infty}
\left(\frac{ x}{l}\right)^2\left.h^{(\infty)}_l
\right|_{\alpha=1-\frac{1}{2}\left(\frac{\mu x}{l}\right)^2}
\end{eqnarray}

\break  \eject

\end{document}